\documentclass[a4paper]{aa}

\usepackage{graphicx}
\usepackage{txfonts}
\usepackage{lipsum}
\usepackage{xcolor}
\usepackage{subcaption}
\usepackage{hyperref}
\usepackage{amsmath}
\usepackage{tabularx, multirow}
\usepackage{siunitx}
\usepackage{soul}
\usepackage{booktabs}

\def\kms{km s$^{-1}$}
\def\kmsmpc{km s$^{-1}$ Mpc$^{-1}$}
\def\cmc{cm$^{-3}$}

\def\ergs{erg s$^{-1}$}

\def\ergscm{erg s$^{-1}$ cm$^{-2}$}
\def\msun{\ifmmode M_{\odot} \else M$_{\odot}$\fi}
\def\msunyr{\ifmmode M_{\odot} {\rm yr}^{-1} \else M$_{\odot}$ yr$^{-1}$\fi}
\def\msunyrvol{\ifmmode \msunyr {\rm Mpc}^{-3} \else \msunyr Mpc$^{-3}$\fi}
\def\zsun{\ifmmode Z_{\odot} \else Z$_{\odot}$\fi}
\def\lsun{\ifmmode L_{\odot} \else L$_{\odot}$\fi}

\newcommand{\balmer}[1]{\ifmmode \text{H}#1 \else H$#1$\fi}
\newcommand{\ha}{\balmer{\alpha}}
\newcommand{\hb}{\balmer{\beta}}
\newcommand{\hg}{\balmer{\gamma}}
\newcommand{\hd}{\balmer{\delta}}
\def\he{\balmer{\epsilon}}

\newcommand{\forbidden}[2]{\ifmmode\textrm{[#1~{\sc #2}]}\else[#1~{\sc #2}]\fi}

\def\oi{\forbidden{O}{i}}
\def\oii{\forbidden{O}{ii}}
\def\oiii{\forbidden{O}{iii}}

\def\nii{\forbidden{N}{ii}}

\def\sii{\forbidden{S}{ii}}
\def\feii{\forbidden{Fe}{ii}}
\def\fevii{\forbidden{Fe}{vii}}

\def\neiii{\forbidden{Ne}{iii}}

\newcommand{\recombination}[2]{\ifmmode\textrm{#1~{\sc #2}}\else#1~{\sc #2}\fi}

\newcommand{\singlet}[2]{\ifmmode{\rm #1}\lambda#2\else #1$\lambda$#2\fi}
\newcommand{\doublet}[3]{\ifmmode{\rm #1}\lambda\lambda#2,#3\else#1$\lambda\lambda$#2,#3\fi}
\newcommand{\triplet}[4]{\ifmmode{\rm #1}\lambda\lambda\lambda#2,#3,#4\else#1$\lambda\lambda\lambda$#2,#3,#4\fi}

\def\Oii{\doublet{\oii}{3727}{3730}}

\def\Oiii{\doublet{\oiii}{4960}{5008}}
\def\Oiiit{\singlet{\oiii}{4960}}
\def\Oiiib{\singlet{\oiii}{5008}}
\def\Oiiia{\singlet{\oiii}{4364}}

\def\Nii{\doublet{\nii}{6550}{6585}}
\def\Niit{\singlet{\nii}{6550}}
\def\Niib{\singlet{\nii}{6585}}

\def\Sii{\doublet{\sii}{6718}{6733}}
\newcommand{\Siis}[1]{\singlet{\sii}{#1}}

\def\ariv{\forbidden{Ar}{iv}}

\newcommand{\Oi}[1]{\singlet{\oi}{#1}}

\newcommand{\Neiii}[1]{\singlet{\neiii}{#1}}

\newcommand{\Feii}[1]{\singlet{\feii}{#1}}
\newcommand{\Fevii}[1]{\singlet{\fevii}{#1}}

\newcommand{\Ariv}[1]{\singlet{\ariv}{#1}}

\def\hei{\recombination{He}{i}}
\def\heii{\recombination{He}{ii}}

\newcommand{\Hei}[1]{\singlet{\hei}{#1}}
\newcommand{\Heii}[1]{\singlet{\heii}{#1}}

\def\oiiihb{\hb+\oiii}
\def\Oiiihb{\hb+\Oiii}

\def\nai{\recombination{Na}{i}}

\def\Nai{\doublet{\nai}{5890}{5896}}
\def\Nait{\singlet{\nai}{5890}}

\def\mgi{\recombination{Mg}{i}}
\def\caii{\recombination{Ca}{ii}}
\def\ki{\recombination{K}{i}}
\def\HeiNA{\Hei{5876}}

\newcommand{\oh}{\ifmmode 12 + \log({\rm O/H}) \else$12 + \log({\rm
O/H})$\fi}

\newcommand{\mstar}{\ifmmode M_\star \else $M_\star$\fi}
\newcommand{\muv}{\ifmmode M_\text{UV}\else $M_\text{UV}$\fi}
\newcommand{\auv}{\ifmmode A_{\rm UV} \else $A_{\rm UV}$\fi}
\newcommand{\luv}{\ifmmode L_{\rm UV} \else $L_{\rm UV}$\fi}
\newcommand{\lir}{\ifmmode L_{\rm IR} \else $L_{\rm IR}$\fi}
\newcommand{\lbol}{\ifmmode L_{\rm bol} \else $L_{\rm bol}$\fi}
\newcommand{\liruv}{\ifmmode L_{\rm IR+UV} \else $L_{\rm IR+UV}$\fi}
\newcommand{\liroveruv}{\ifmmode L_{\rm IR}/L_{\rm UV} \else $L_{\rm IR}/L_{\rm UV}$\fi}
\newcommand{\nlyc}{\ifmmode N_{\rm Lyc} \else $N_{\rm Lyc} $\fi}
\newcommand{\rholyc}{\ifmmode \rho_{\rm Lyc} \else $\rho_{\rm Lyc} $\fi}
\newcommand{\chion}{\ifmmode \xi_{\rm ion} \else $\xi_{\rm ion}$\fi}
\newcommand{\chioncorr}{\ifmmode \xi_{\rm ion}^0 \else $\xi_{\rm ion}^0$\fi}
\newcommand{\Rthree}{\ifmmode R3 \else $R3$ \fi}
\newcommand{\Rthreefunc}{\ifmmode \Rthree(\muv) \else $\Rthree(\muv)$ \fi}
\def\BHM{\ifmmode M_{\rm BH}\else $M_{\rm BH}$\fi}

\newcommand{\fesc}{\ifmmode f_\textrm{esc} \else $f_\textrm{esc}$ \fi}
\newcommand{\abs}[1]{\left|#1\right|}
\def\bhs{BH$^*$}

\newcommand{\on}{GLIMPSED-329380}
\newcommand{\mlw}{\ifmmode M_\text{LW}\else $M_\text{LW}$\fi}

\begin{document} 
    \title{GLIMPSED: Direct evidence for a fast active galactic nucleus-driven outflow from a $z=6.64$ little red dot host galaxy}

    \titlerunning{A fast AGN-driven outflow from a $z=6.64$ LRD host galaxy}

   \author{Damien Korber\inst{1}\thanks{Corresponding author: Damien Korber:\\ \href{mailto:damien.korber@protonmail.ch}{damien.korber@protonmail.ch}}
          \and Rui Marques-Chaves\inst{1}
          \and Daniel Schaerer\inst{1,2}
          \and Gabriel Brammer\inst{3}
          \and Archana Aravindan\inst{4,5}
          \and Arghyadeep Basu\inst{6}
          \and Qinyue Fei\inst{7}
          \and Emma Giovinazzo\inst{1}
          \and Vasily Kokorev\inst{4}
          \and Alberto Saldana-Lopez\inst{8}
          \and Maxime Trebitsch\inst{9}
          \and Hakim Atek\inst{10}
          \and John Chisholm\inst{4,5}
          \and Ryan Endsley\inst{4}
          \and Seiji Fujimoto\inst{7,11}
          \and Lukas Furtak\inst{4,5}
          \and Richard Pan\inst{12}
          \and Rohan P. Naidu\inst{13}
          }
    
   \institute{Observatoire de Genève, Université de Genève, Chemin Pegasi 51, 1290 Versoix, Switzerland
        \and CNRS, IRAP, 14 Avenue E. Belin, 31400 Toulouse, France
        \and Cosmic Dawn Center (DAWN), Niels Bohr Institute, University of Copenhagen, Jagtvej 128, København N, DK-2200, Denmark
        \and Department of Astronomy, The University of Texas at Austin, Austin, TX 78712, USA
        \and Cosmic Frontier Center, The University of Texas at Austin, Austin, TX 78712, USA
        \and Univ Lyon, Univ Lyon1, Ens de Lyon, CNRS, CRAL UMR5574, F-69230, Saint-Genis-Laval, France
        \and David A. Dunlap Department of Astronomy and Astrophysics, University of Toronto, 50 St. George Street, Toronto, Ontario, M5S 3H4, Canada
        \and Department of Astronomy, Oskar Klein Centre, Stockholm University, 106 91 Stockholm, Sweden
        \and LUX, Observatoire de Paris, Université PSL, Sorbonne Université, CNRS, 75014 Paris, France
        \and Institut d'Astrophysique de Paris, UMR 7095, CNRS, Sorbonne Université, 98 bis boulevard Arago, 75014 Paris, France
        \and Dunlap Institute for Astronomy and Astrophysics, University of Toronto, Toronto, ON M5S 3H4, Canada
        \and Department of Physics \& Astronomy, Tufts University, MA 02155, USA20
        \and MIT Kavli Institute for Astrophysics and Space Research, 70 Vassar Street, Cambridge, MA 02139, USA
        }

   \date{Received 09 April 2026; accepted 11 September 2026}

    \abstract{We report the discovery of \on, a $z=6.64$ galaxy behind Abell S1063, which shows signs of an extreme ionised outflow driven by an active galactic nucleus (AGN). The deep JWST/NIRSpec medium grating observations show spatially resolved structures of a host galaxy containing the very fast outflow and an AGN, which we analyse separately. The outflow, mainly traced by broad \Oiii\ and \ha\ emissions in the host, reaches a full-width-at-half-maximum velocity of $\sim5350$\kms, velocities only observed in AGN-dominated systems. From the Balmer decrement, we observe that while the narrow emission lines show no dust attenuation, the outflowing gas is dusty. We use emission lines diagnostics to infer gas abundances within the host galaxy. The oxygen abundance is $12+\log({\rm O/H}) \sim 7.96$ ($\sim 18\%$ solar) and the host is slightly nitrogen-enriched with $\log({\rm N/O})\sim -0.68$.
    Unless the extreme outflow turns out to be very dusty, the mass loading factor, $\eta > 5-10\%$, and the kinetic energy, $>10^{43}$\ergs\ , of the extreme outflow does not hint at a significant impact on the galaxy. The AGN component shows many similarities with little red dots (LRDs): a characteristic ‘V shape’, exponential profile in hydrogen lines, numerous detections of forbidden \feii\ lines, a Balmer break, and a broad absorption feature at $\sim4550$ Å. This detection of a fast outflow in an LRD, rare in surveys dominated by low-resolution (e.g. PRISM) spectra, provides direct evidence of AGN activity in these systems.}

   \maketitle
 \nolinenumbers

\section{Introduction}

Outflows are a fundamental mechanism for regulating galaxy evolution, driven by processes acting across different scales. On local scales, starburst activity powers outflows with velocities of hundreds of kilometres per second \citep[e.g.][]{Lynds_evidence_1963, Heckman_nature_1990, Veilleux_galactic_2005, Rupke_outflows_2005, arribas_ionized_2014, forster_schreiber_kmos3d_2019, roberts-borsani_outflows_2020, llerena_ionized_2023, saldana-lopez_feedback_2025, rodriguez_del_pino_ga-nifs_2026}, shaping the interstellar medium and disturbing star formation \citep[SF; e.g.][]{hopkins_galaxies_2014}. In the early Universe, such feedback may contribute to the bursty SF histories observed in high-redshift galaxies \citep[e.g.][]{endsley_star-forming_2024}. On larger scales, accreting supermassive black holes in active galactic nuclei (AGNs) drive far more energetic outflows, reaching velocities of up to several thousand kilometres per second \citep[e.g.][]{Silk_quasars_1998,  Veilleux_galactic_2005, arribas_ionized_2014, king_powerful_2015, forster_schreiber_kmos3d_2019, bertola_ga-nifs_2025, nandi_stratification_2026}. AGN-driven outflows can disrupt the galaxy gas reservoir on a larger scale, potentially quenching SF \citep[e.g.][]{forster_schreiber_kmos3d_2019, veilleux_cool_2020, cooper_high-velocity_2025}. By tracing the kinematics and physical conditions of outflowing gas, we gain insights into the energy sources powering them and their impact on galaxy growth.

The James Webb Space Telescope has revealed in recent years a new category of objects called little red dots \citep[LRDs, ][]{matthee_little_2024}. These objects were identified as red, unresolved dots in JWST/NIRCam observations, showing a V-shaped SED and broad Balmer emission lines \citep[e.g.][]{matthee_little_2024, furtak_high_2024, greene_uncover_2024}. These objects have gained a lot of attention over the past years and have been studied in many different ways, revealing common properties. These objects potentially contain high-density gas envelopes, as shown by the presence of exponential wings, non-stellar Balmer breaks, and frequent Balmer absorption features \citep[e.g.][]{Labbe_unambiguous_2024, Wang_rubies_2024, juodvbalis_jades_2024, ji_blackthunder_2025, deugenio_irony_2025, lambrides_discovery_2025, ji_lord_2026, lin_discovery_2026, rusakov_little_2026}. Furthermore, deeper observations or LRD stacks often reveal the presence of forbidden \feii\ and permitted Fe {\sc ii} lines \citep[e.g.][]{Labbe_unambiguous_2024, Tripodi_deep_2025, lambrides_discovery_2025, deugenio_irony_2025, Torralba_warm_2026, Kokorev_deepest_2026, lin_discovery_2026, perez-gonzalez_little_2026}. Their origin remains unclear to date, with some authors suggesting the presence of a black hole with a nearby dense gas clump \citep[][]{inayoshi_extremely_2025} or a black hole surrounded by a gas envelope with large covering factor \citep[e.g. various approaches:][]{kido_black_2025, Naidu_black_2026, Kokorev_deepest_2026, Begelman_little_2026, degraaff_little_2026}. This gas would be heated by the black hole, which then produces radiation, and hence the observed black-body shaped SED \citep[e.g.][]{degraaff_little_2026, sun_little_2026}.

Outflows in LRDs are interesting, because while low-redshift AGN populations lack the neutral gas absorption that is commonly observed in the Balmer and helium lines of LRDs \citep[e.g.][]{juodvbalis_jades_2024, matthee_little_2024, Wang_rubies_2024, wang_rubies_2025, Kocevski_rise_2025, Torralba_warm_2026, degraaff_little_2026, DEugenio_BlackTHUNDER_2026, Naidu_black_2026, deugenio_jades_2026}, LRDs often lack ionised outflows which are observed in low-redshift AGNs \citep[e.g.][]{arribas_ionized_2014, forster_schreiber_kmos3d_2019, bertola_ga-nifs_2025}. Nonetheless, a growing number of studies report the presence of ionised outflows in LRDs. \citet{cooper_high-velocity_2025} identified an ionised outflow in a $z\sim7.04$ LRD from \citet{rinaldi_not_2025}. In their sample, this object exhibits the fastest outflow at $z>5$, with velocities reaching $v_{\rm out}\sim758$\kms\ (see App.~\ref{app:vout} for our definition of $v_{\rm out}$). The authors report a companion for this LRD, which could then enhance the outflow due to merging activities \citep{cooper_high-velocity_2025}. \citet{deugenio_jades_2026} identified a broad but relatively faint component in \Oiiib\ reaching $v_{\rm out}\sim 269$\kms~in a $z\sim5$ LRD.\cite{juodvbalis_jades_2024} reported a weak ionised outflow in a $z\sim2$ LRD. The velocity in \oiii\ is $v_{\rm FWHM}\sim1310$\kms, and the outflow is slightly redshifted. At lower redshift ($z\sim0.4)$, \citet{chen_z_2026} reported an extreme ionised outflow reaching $v_{\rm out}\sim 3097$\kms, in an LRD-like galaxy. Finally, \citet{wang_rubies_2025} reported an ionised outflow of $\sim234$\kms\ in an LRD at $z=3.1$, through an absorption feature in \singlet{\hei}{1.083$\mu$m}. They mention that this outflowing velocity matches star-formation driven outflows at $z\sim2$ in \citet{forster_schreiber_kmos3d_2019}, but their estimated physical scale is much lower \citep{wang_rubies_2025}.

Intuitively, if LRDs host actively accreting black holes, one might expect feedback-driven outflows. However, such outflows are only rarely observed. Recent scenarios propose that a dense gas envelope with large covering surrounding the black hole could explain this apparent discrepancy \citep{inayoshi_extremely_2025, kido_black_2025, taylor_casper-lrd-z9_2025, Naidu_black_2026, Kokorev_deepest_2026, Begelman_little_2026, degraaff_little_2026}. In these models, gas accelerated by accretion is confined within the envelope, and its energy is reprocessed into radiation \citep{kido_black_2025} or is convectively transported outwards through the envelope before being radiated at the photosphere \citep{Begelman_formation_2006, Begelman_little_2026, Roman-Garza_quasi-star_2026}, producing the observed black-body spectrum \citep{Naidu_black_2026, Begelman_little_2026, degraaff_little_2026}. Hence, outflows can only emerge if the envelope becomes unstable, allowing some gas or radiation to escape.

\begin{figure*}[h]
    \centering
    \includegraphics[width=\textwidth]{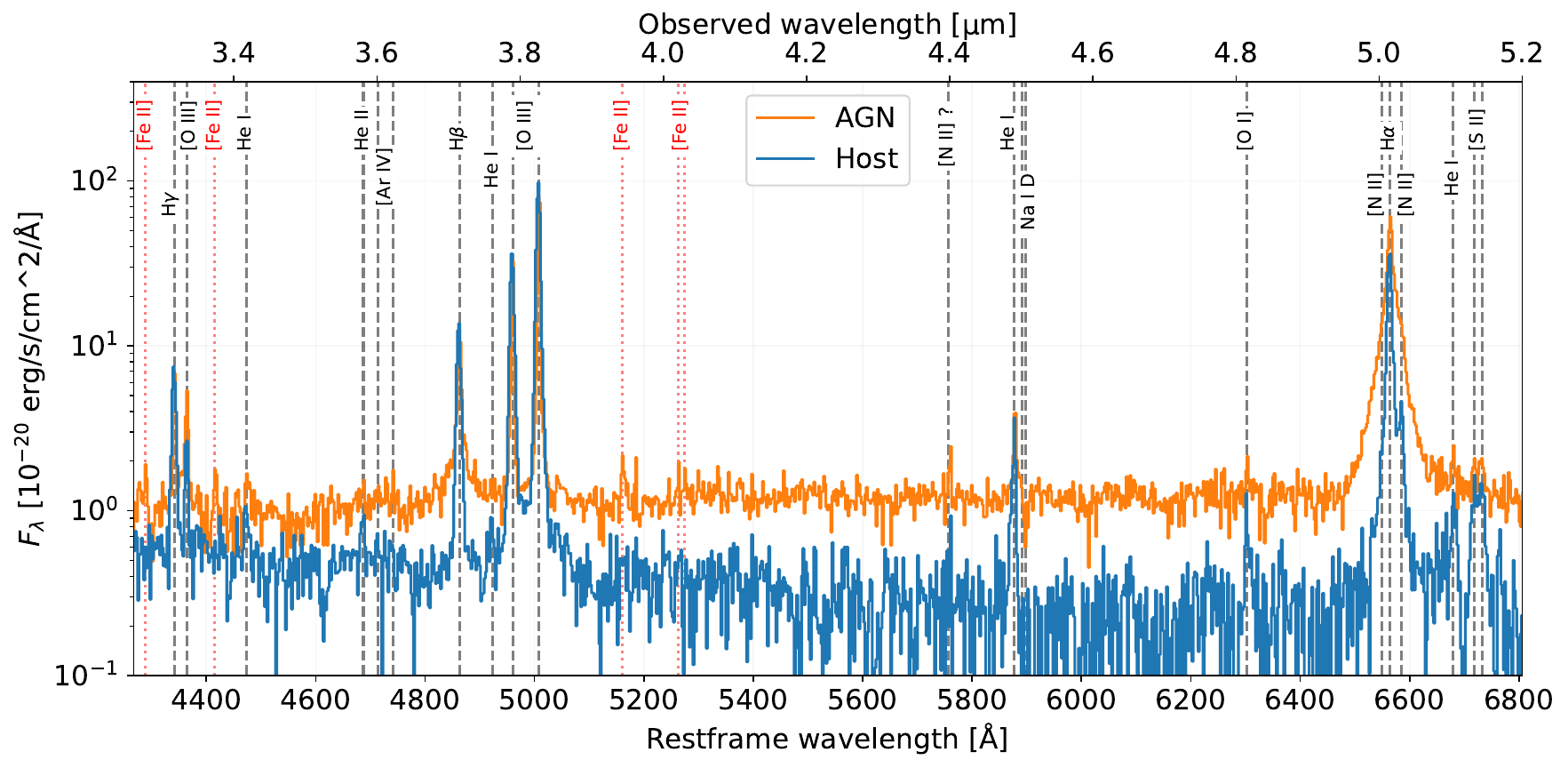}
    \caption{Separated spectra of the AGN (orange) and host (blue) components with the global noise reduction. The spectrum is restricted to highlight the strong lines.}
    \label{fig:reduction}
\end{figure*}

In this paper, we present \on, a $z=6.64$ galaxy showing signs of an extreme outflow, alongside a compact AGN exhibiting typical LRD features. This galaxy was initially observed as a filler of the deep medium band GLIMPSED survey \citep[DDT-9223:][]{fujimoto_glimpsed_2025}. We present this emission-line-rich spectrum, which can be separated into two components due to being resolved in the 2D spectrum (Sect. \ref{sec:spectroscopy}) as well as the limited photometric observations (Sect. \ref{sec:photometry}). We then detail the fitting methods (Sect. \ref{sec:fit_host} and \ref{sec:fit_agn}). The results are then presented separately for the two components, starting with the host component, which presents the extreme outflow (Sect. \ref{sec:res_host}), and then the AGN component, describing the different features visible in the spectrum (Sect. \ref{sec:res_agn}). Finally, we discuss the different LRD-like features, which tell us this object is an LRD (Sect. \ref{sec:disc_lrd}), then we discuss the outflow in the context of LRDs (Sect. \ref{sec:nature_outflow_lrd}). We considered a flat $\Lambda$ cold dark matter cosmology with $H_0 = 70$ \kmsmpc, $\Omega_M = 0.3$, and $\Omega_\Lambda = 0.7$.

\section{Data and line measurements}\label{sec:data}

\subsection{Spectroscopy}\label{sec:spectroscopy}

\on\ lies on the outskirts of the Abell S1063 galaxy cluster (RA=342.229976deg, Dec=-44.510380deg). This galaxy was observed for 9 hrs with the G395M medium-resolution spectroscopy of JWST/NIRSpec, as part of the GLIMPSED survey \citep[DDT-9223; ][]{fujimoto_glimpsed_2025}.
We initially started using the reduction from the DJA\footnote{\url{https://dawn-cph.github.io/dja/spectroscopy/nirspec/}} \citep{Heintz_strong_2024, degraaff_rubies_2025}, which was produced using \texttt{msaexp}, following the standard JWST reduction pipeline \citep{brammer_msaexp_2023}.
In general, the approach to remove the diffuse sky component from the MSA spectra is to take 2D image differences of exposures taken with spatial `nod' offsets of the telescope. With the small 5-pixel ($\sim$0\farcs5) nods necessary to keep the target in the three-shutter MSA slitlet in all exposures, this can result in `self-subtraction' in the outer profiles of extended sources\footnote{The variance of the differenced exposures is also increased by a factor of $1/N_b$, where $N_b$ is the number of exposures used to define the source-free `sky' and $N_b=8$ here.}. 
Then, for all the slitlets in the MSA configuration, \texttt{msaexp} estimates the 1D spectrum of the sky of an exposure using the portions of the spectra excluding the expected traces of the sources within the slitlets. \on\ is resolved in NIRSpec, and suffers from such self-subtraction in strong emission lines (e.g. \oiii\ and \ha).

To bypass this issue, we instead adopted a `global' sky subtraction approach by foregoing the nod differences and subtracting this 1D sky spectrum from the slit 2D spectra before the combination and rectification of the final spectrum. This technique offers better sensitivity and suppresses the self-subtraction issues from the nod difference method. We redirect the readers to Sect.~3 and App.~A.2 in \citet{degraaff_rubies_2025}, which describes this global sky subtraction method in great detail.

Finally, the 1D spectrum was extracted from the 2D spectrum using the standard \texttt{msaexp} pipeline. Almost all the results of this paper use the global sky background subtraction. However, due to artefacts in the continuum in the blue edge of the spectrum ($\lesssim3.25\mu$m), the few measurements in the continuum used the nod difference sky subtraction instead.

\begin{figure}[t]
    \centering
    \includegraphics[width=0.9\linewidth]{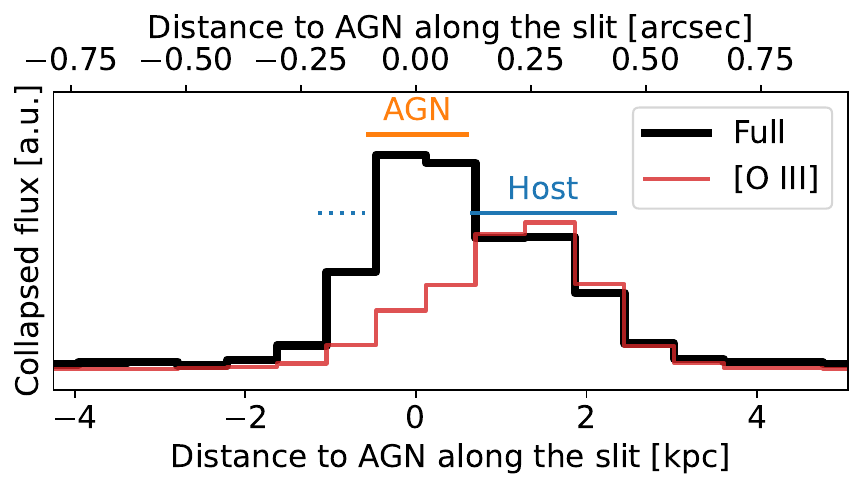}
    \includegraphics[width=1.0\linewidth]{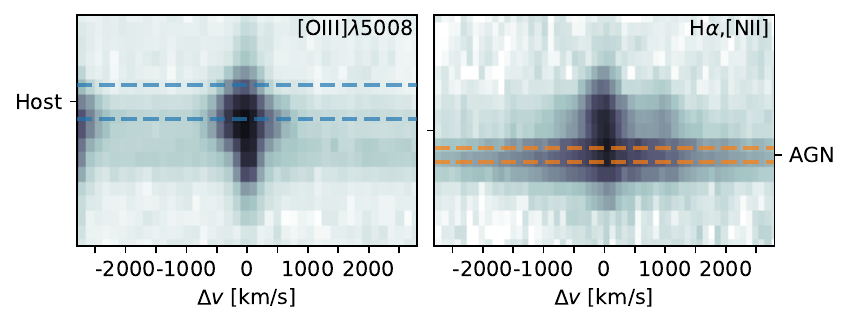}
    \caption{Top panel: Total flux along the wavelength axis for the full spectrum and the \oiii\ region separately. The x axis shows the spatial distance to the AGN on the slit. Bottom panel: 2D spectrum showing the spatial distribution (y axis) of the velocity profiles (x axis) from \Oiiib\ and \ha. The FWHM of the Gaussian profiles used to drizzle the spectra are shown in blue (Host, on \Oiiib) and orange (AGN, on \ha).}
    \label{fig:2d_profile}
\end{figure}

The spectrum of \on\ reveals two spatially distinct components. The first one, hereafter referred to as the AGN component, features a well-detected compact continuum confined between two pixels and exhibits broad \ha\ emission. The second one, the host component, displays a fainter but detected extended continuum, alongside broadened \oiii\ emission in the extended regions. We extracted both components separately using the standard drizzled \texttt{msaexp} extraction \citep{brammer_msaexp_2023}. The flux was extracted with a Gaussian of mean $\mu$ and standard deviation $\sigma$. In both spectra, we manually fixed $\mu$ (to $-1.7$px from centre for the AGN, and to $+2$px from centre for the host) and $\sigma$ (to $0.4$px for the AGN, and to $1$px for the host). These parameters separate the spectrum into two distinct regions to minimise contaminations. The two separate spectra can be visualised in Fig. \ref{fig:reduction}, alongside identified emission lines. In addition, we show in Fig. \ref{fig:2d_profile} the collapsed spectrum (the sum along the frequency axis), illustrating the extent of the galaxy in the JWST/NIRSpec slit, and the 2D spectrum focused around \Oiiib\ with its outflowing component on top and \ha\ with its broad-line region (BLR) and broad component. The full 2D spectrum shows the unresolved AGN, as well as wings corresponding to the contribution of the host. The restricted sum around \oiii\ is spatially offset from the AGN, and shows that oxygen lines spread over several kiloparsecs.

\begin{figure*}[t]
    \centering
    \includegraphics[width=\linewidth]{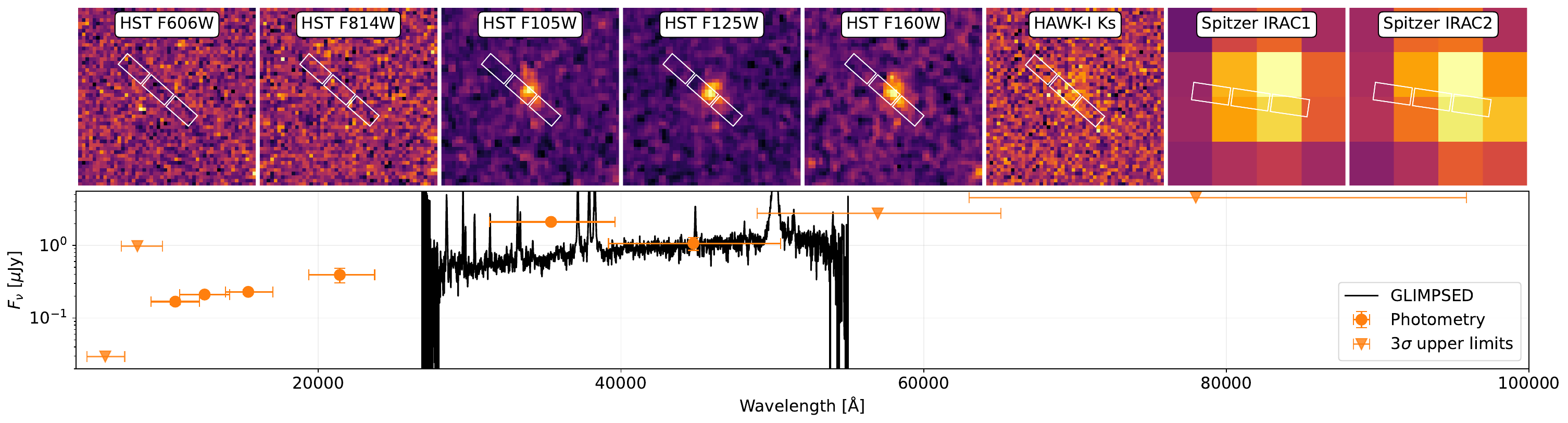}
    \caption{Top panel: Cutout of the available photometry. \on\ is not detected in F606W and F814W due to the redshift, as well as in IRAC3 and IRAC4 (not displayed for readability reason), but is detected in the other filters. The cutouts have square sides of 3 arcsec, and the MSA slit is displayed on top of all the images in white. Bottom panel: Comparison between the available photometry and the full GLIMPSED spectrum (not corrected for slit loss)}
    \label{fig:phot_cutout}
\end{figure*}

\subsection{Photometry}\label{sec:photometry}

While \on\ is located near the Abell S1063 galaxy cluster, it lies just outside the photometric footprint of the ultra-deep GLIMPSE survey \citep[GO-3293;][]{atek_jwst_2025} and was not covered by other JWST surveys. However, the galaxy was detected in several earlier photometric surveys. From space, the BUFFALO HST survey \citep[HST-15117;][]{steinhardt_buffalo_2020} observed this galaxy in five HST/WFC3 rest-frame optical filters (F606W, F814W, F105W, F125W, and F160W) with a long exposure time (16hrs to 67hrs, F814W being the deepest). As F606W and F814W are beyond the Lyman break, the galaxy is only observed in the three red filters. It is also observed in Spitzer/IRAC1-4, but is only detected in IRAC1 and 2 (PID-83). From the ground, the KIFF survey \citep{brammer_ultra-deep_2016} detected it with the VLT/HAWK-I Ks filter with a long exposure time (22hrs). Interestingly, the non-detections in IRAC3 and IRAC4 suggest that the AGN is not heavily obscured, which would otherwise be more prominently detected at longer wavelengths. To our knowledge, these are all the available photometric detections, which significantly limits its morphological study. Cutouts of this galaxy are displayed in Fig. \ref{fig:phot_cutout}.

\subsection{Modelling the host component}\label{sec:fit_host}

The host component is particularly interesting for the presence of strong outflows.
In order to characterise the host and its outflows, we fitted the object in four regions of interest: the regions surrounding  \Oii, \hg+\Oiiia, \hb+\Oiii\ , and \ha+\Nii+\Sii.

\begin{figure*}
    \centering
    \includegraphics[width=\linewidth]{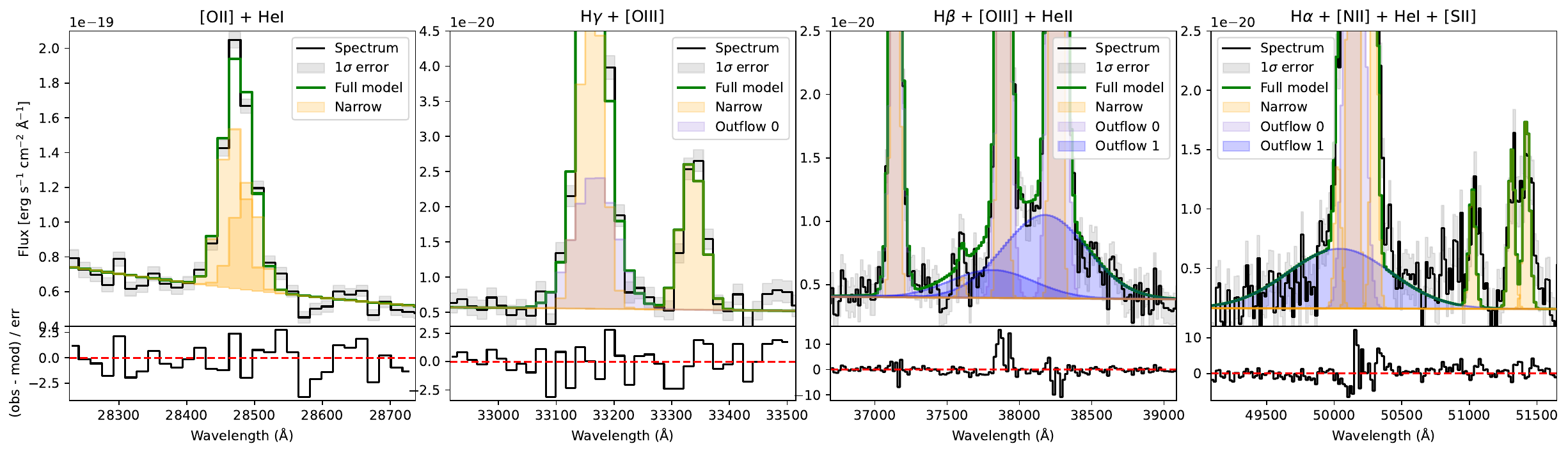}
    \caption{Top panel: Best fit of the strong emission lines in the host, with the narrow and the broad components. Bottom panel: Residual of the total fit (green) and the data (black). A detailed view of \oiiihb\ is displayed in Fig.~\ref{fig:outflow_vs_no}.}
    \label{fig:host_el_fit}
\end{figure*}

We first fitted the continuum in each spectral region using a linear fit of the surrounding continuum, except for \oii, for which we used a second-degree polynomial due to increased systematic effects at the spectral edges (note that \oii\ lie in the extended region of the spectrum). We subtracted the continuum from the spectrum to more easily fit the emission lines.
We then separately modelled each emission line using one narrow Gaussian, and for some broader lines, one or two outflow broad components (each with three parameters: the amplitude, $A_i$, the standard deviation, $\sigma_i$, and the mean, $\mu_i$).
To limit the number of fit parameters, we used several physical considerations. The position of each line is fixed to the rest-frame wavelength, and is redshifted using the systemic redshift, $z_{\rm}$. For each broad component (called outflow 0 and 1 in the following), we chose one common redshift respectively. We therefore reduced the number of emission line wavelength parameters to three redshift parameters. 
The line width of each component was fixed to a common intrinsic FWHM (in kilometres per second). The latter was convolved with the grating resolving power (see App.~\ref{app:table_emission_lines}). This reduced the number of width parameters to three as well. Finally, the amplitude of each emission line was kept free, unless fixed by atomic physics. This is the case for $\Oiiib / \Oiiit = 2.98$ \citep{storey_theoretical_2000} and $\Niib / \Niit = 2.96$ \citep{storey_theoretical_2000}, for which only one amplitude is free. We then used \texttt{LMFIT} \citep{newville_lmfit_2025} to optimise the parameters. 
The measured line fluxes of the narrow and broad components (if present) of the host are reported in Table \ref{tab:host_emission_lines}.
The emission line fits can be visualised in Fig. \ref{fig:host_el_fit}. A comparison of the fit with and without outflows is shown in App.~\ref{app:stat_ana_outflow}, and we define $\Delta{\rm BIC} > 10$ as statistically significant for the rest of the paper.

\subsection{Modelling the AGN component}\label{sec:fit_agn}

For the AGN component, we report all the emission lines observed in the spectrum. Due to the large number of emission lines observed in that component, we opted for a semi-automated modelling of the continuum, with manual selection of the fitting regions for strong emission lines. We first excluded a region of $300$Å around the emission lines considered in the model (as tabulated in Tables \ref{tab:agn_emission_lines} and \ref{tab:agn_iron_forest}). In these regions, we assumed the continuum to be linear. As broad lines span regions larger than $300$Å, we enforced a linear continuum as well. These regions are \Oii, \hd, \hg+\Oiiia, \hb+\Oiii, \singlet{\hei}{5877}+\Nai\ , and \ha+\Nii+\Sii. We then used the median filter from \texttt{Scipy} \citep{virtanen_scipy_2020}, with a kernel size of 51 pixels, to obtain an estimation of the continuum, which was subtracted from the spectrum to improve the fitting (see App.~\ref{app:agn_continuum}).

\begin{figure*}
    \centering
    \includegraphics[width=1\linewidth]{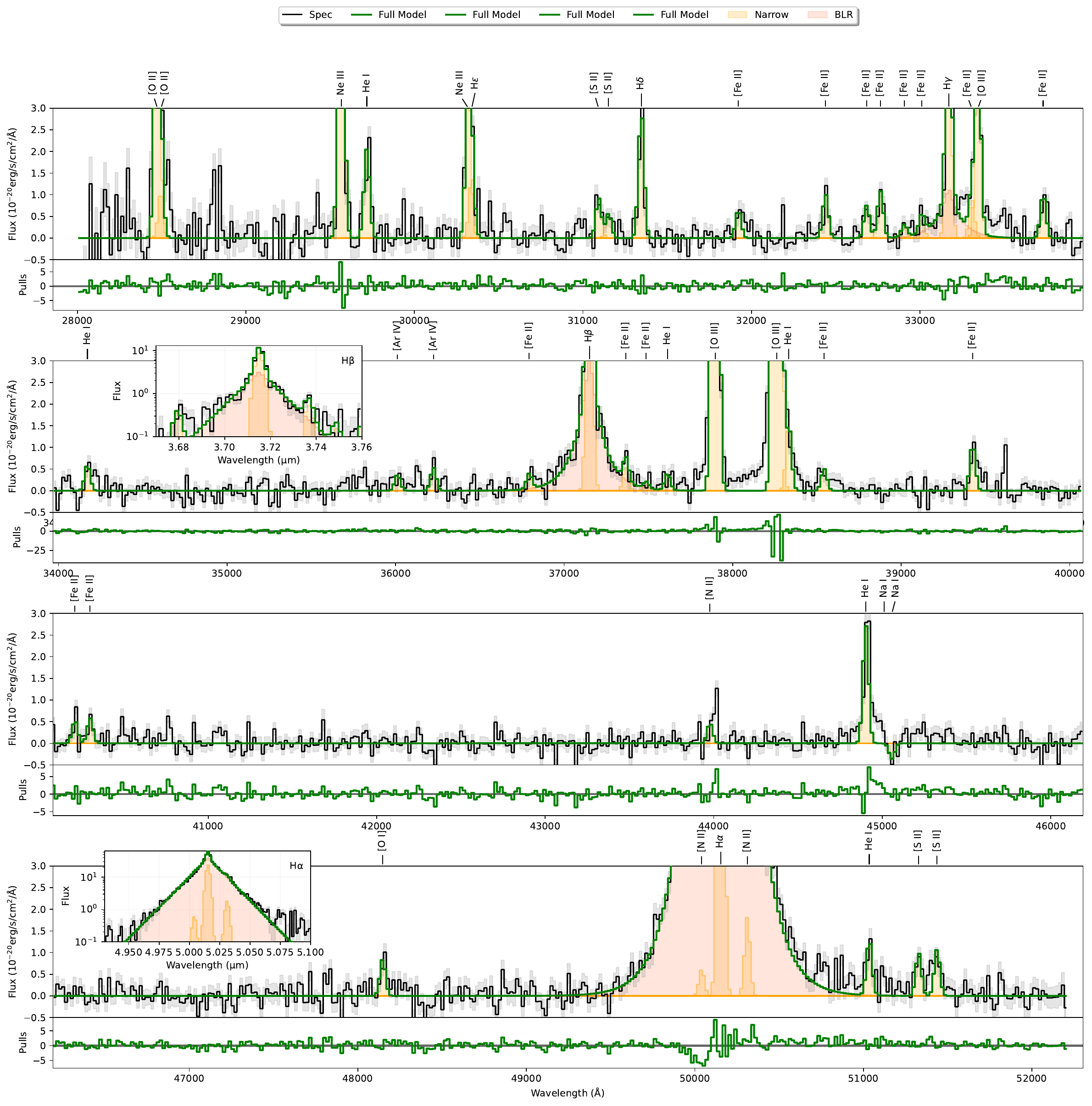}
    \caption{Background-subtracted spectrum of the AGN component. This figure includes the fit of emission lines that are found in Table \ref{tab:agn_emission_lines}. Zoomed-in boxes of \ha\ and \hb\ are displayed on a logarithmic scale with the same flux units. We also display the Pulls = (data - model)/uncertainty at the bottom of each spectrum.}
    \label{fig:agn_spec_fig}
\end{figure*}

We fitted the emission lines in a similar fashion to the host galaxy. It only differs in the broad components. After testing different configurations, we discarded any outflowing component in oxygen (and Balmer) lines. The reason for this is that contaminations from the host perturb the wings, and produce complex emission line profiles that are not well reproduced by a combination of Gaussians. 
In the Balmer lines, we consider a line broadening from the BLR. In order to fit the BLR wings, we tried to fit both a Gaussian and an exponential. We observe a slightly better fit with the exponential model ($\Delta{\rm BIC} \sim 59$), so we use it for the final fit. The redshift of the BLR is fixed to the systemics, as we do not expect variation in redshift. All the lines considered for the AGN are reported in Table \ref{tab:agn_emission_lines}, except for forbidden iron lines reported in Table \ref{tab:agn_iron_forest}. All the fits can be visualised in Fig.~\ref{fig:agn_spec_fig}.

\section{Results}\label{sec:results}

\subsection{Host galaxy properties}\label{sec:res_host}

Using the method described earlier, we find that most of our emission lines are well fitted by a single narrow line component. For the strongest lines (\hg, \hb, \Oiii\ , and \ha), the addition of two broad components strongly improves the fit of the wings. The first one has ${\rm v_{\rm FWHM}\sim740-787}$\kms\ and the second shows much larger wings, ${\rm v_{\rm FWHM}\sim5345-5349}$\kms.
These broad components can be visualised in Fig. \ref{fig:host_el_fit}. The latter is extreme and is unlikely to be related to the virial motion of the gas. It is consistent with a strong ionised outflow driven by an AGN. Therefore, from now on we consider these broad components to represent ionised outflowing gas.

\subsubsection{Dust attenuation in the outflow}

In Table \ref{tab:host_emission_lines}, we report significant detection of outflows in successive Balmer series emission lines. 
In the absence of dust, galaxies are expected to follow ratios of $\ha/\hb=2.86$ and $\hb/\hg = 2.13$, assuming case B recombination and $T_e=10000$K \citep{osterbrock_astrophysics_2006}. Our narrow line analysis yields a ratio of $\ha/\hb = 3.09 \pm 0.24$ and $\hb/\hg = 2.04 \pm 0.28$. Then, the first outflow component yields $\ha/\hb = 3.59 \pm 0.70$ and $\hb/\hg = 1.91 \pm 0.54$. Unfortunately, in the faster outflow, \hb\ is unconstrained; therefore, we only constrain to a lower limit $\ha/\hb > 3.68$.
The narrow component is thus compatible with case B recombination, indicating minimal dust attenuation. In contrast, the outflows show tentative evidence of dust. The Balmer decrement ($\ha/\hb$) of the first outflow slightly deviates from case B, indicating significant attenuation. For the second outflow, only a lower limit is measured, which is already redder than case B.
We note a slight inconsistency in the attenuation inferred with the second tracer ($\hb/\hg$) in the first outflow, as it suggests no decrement. Nonetheless, due to large uncertainties on both \hb\ and \hg, the ratios are compatible with minimal attenuation.
In this picture, we expect the dust within the host galaxy to be ejected alongside the outflowing ionised gas, which matches recent observations of dusty outflows \citep{crespo-gomez_rioja_2025, Marques-Chaves_extremely_2026, rodriguez_del_pino_ga-nifs_2026}.
In the rest of the results, the dust-corrected emission lines from the host component will assume the \citet{cardelli_relationship_1989} extinction curve with an attenuation of $E(B-V)_l = 0.23 \pm 0.20$ for the first outflow and $E(B-V)_l >0.26$ for the fast outflow. The narrow lines are not corrected for dust.

\subsubsection{Electron temperature, electron density, and abundances of the host} \label{sec:te_ne}

The electron temperature, $T_e$, and density, $n_e$, can be estimated using ratios of emission lines sensitive to these quantities.
To estimate $T_e$, we can use the direct method from the ratio of \Oiiia and \Oiiib\ \citep{Sanders_aurora_2026}, and for $n_e$, we use the \doublet{\sii}{6718}{6733} doublet \citep{kewley_theoretical_2019}. 
In \on, the \doublet{\sii}{6718}{6733}\ doublet is detected, but seems to be affected by the outflow (as is seen in the wings of \sii\ in Fig.~\ref{fig:host_el_fit}). Unfortunately, due to the weak signal, the outflow in \sii\ is not well detected, and the best fit is worse with an outflow component. Therefore, we only assume a narrow component. As the $n_e$ and $T_e$ diagnostics are interdependent, we need to fit them together. We do this with \texttt{PyNeb} \citep{luridiana_pyneb_2015}, finding an 
electron temperature of $T_e = 16521_{-924}^{+845}$ K and density of $n_e = 1745_{-1040}^{+2293}$ \cmc. The final measurements are found in Table \ref{tab:general_properties}.

\begin{table}[t]
\tiny
    \centering
    \caption{Measured properties of \on.}
    \begin{tabularx}{\columnwidth}{lXl}
    \toprule
    Property & Units & Value\\
    \midrule
    \multicolumn{3}{c}{General properties}\\
    RA  & deg & 342.22997617 \\
    DEC & deg & -44.5103798  \\
    $\mu$ & --- & $2.36\pm0.01$  \\
    &&\\
    \multicolumn{3}{c}{host component}\\
    
    $z_{\rm sys}^{\rm host}$ & --- & $6.63785\pm0.00004$\\
    $T_e(O^{+2})$ & K & $16521_{-924}^{+845}$\\
    $n_e(S^{+})$ & \cmc & $1745_{-1040}^{+2293}$\\
    $12+\log({\rm O/H})$ & --- & $7.96 \pm 0.06$\\
    $\log({\rm N/H})$ & --- & $-5.42 \pm 0.05$ \\
    $\log({\rm N/O})$ & --- & $-0.68 \pm 0.10$ \\
    $\log(\mstar)^\S$ & \msun & $8.21 \pm 0.23$\\
    &&\\
    \multicolumn{3}{c}{AGN component}\\
    $z_{\rm sys}^{\rm agn}$ & --- & $6.63994 \pm 0.00003$\\
    $T_e(O^{+2})$ & K & $24357_{-940}^{+484}$\\
    $\log n_e(S^+)$ & \cmc & $3.35_{-0.56}^{+0.50}$\\
    $\log\BHM$ $^\dagger$ & \msun & $6.96 \pm 0.50$\\
    &&\\
    \multicolumn{3}{c}{Supplementary properties}\\
    $\Delta v_{\rm sys}$ & \kms & $-82.23 \pm 1.79$\\
    \muv$^\ddagger$ & --- & $-20.40 \pm 0.07$\\
    $\beta_{\rm UV}$ & --- & $-1.60 \pm 0.04$\\
    $\beta_{\rm opt}$ & --- & $0.37 \pm 0.05$\\
    \bottomrule\end{tabularx}
    \tablefoot{$^\dagger$ The \BHM\ was measured in \ha\ from an exponential profile, so it might be underestimated (see Sect. \ref{sec:bhm}). 
    $^\S$ The stellar mass was indirectly measured from the mass-metallicity relationship of \citep{chemerynska_extreme_2024}. $^\ddagger$ \muv\ was estimated by interpolating the flux at 1.5$\mu$m from the F125W and F160W HST filters. Masses and magnitudes are de-lensed.}
    \label{tab:general_properties}
\end{table}

We further measured the abundance of the main metals in our galaxy. We used \texttt{PyNeb} to infer these quantities. We measured the oxygen abundance with the direct method using \Oiiib\ and \Oii. We find an oxygen abundance of $12+\log({\rm O/H}) = 7.96 \pm 0.06$. We also detect \Nii; therefore, we converted the $T_e(O^{+2})$ to $T_e(O^+)$ from the intermediate $Z$ relationship from \citet{izotov_chemical_2006} and inferred a nitrogen abundance of $\log({\rm N/H}) = -5.42 \pm 0.05$, assuming $T_e(N^+) = T_e(O^+)$ \citep[e.g.][]{berg_characterizing_2021}. By comparing \Oii\ and \Nii, we infer $\log({\rm N/O}) = -0.68 \pm 0.10$. This N/O abundance is slightly higher than typical standard star-forming galaxies, but does not reach values observed in N-enhanced galaxies recently observed at high redshift \citep[e.g.][]{Cameron_nitrogen_2023, marques-chaves_extreme_2024, Schaerer_discovery_2024, Topping_deep_2025, Navarro-Carrera_interstellar_2025, Morel_discovery_2026, berg_fleeting_2026}. Nonetheless, N-enhancement matches observations that broad-line AGNs are often N-enriched at high z \citep{Isobe_jades_2025}.
Because of the difficulties in systematically constraining all the outflowing components, in particular for weaker lines, measuring the same quantities for the outflowing gas remains challenging.

\subsubsection{Extreme ionised outflow}

One of the main points of interest of \on\ is the extreme ionised outflow observed in strong emission lines, in particular \oiii. This component is observed in Fig. \ref{fig:host_el_fit}. As shown in Tab. \ref{tab:host_emission_lines}, the broad components in \oiii\ have a significant full width at half maximum (FWHM) of the order of $\sim$770 \kms\ and $\sim$5350 \kms. In addition, the outflows are visibly shifted from the systemic redshift, with shifts of $\Delta v\sim-58$\kms\ and $\Delta v\sim594$\kms\ , respectively. The latter is extreme and is rarely seen for ionised outflows \citep[e.g.][]{arribas_ionized_2014, forster_schreiber_kmos3d_2019, cooper_high-velocity_2025, nandi_stratification_2026}. The first broad component is fitted at the same redshift as the narrow component, but the broader one appears slightly blueshifted. Outflowing broad components with ${\rm FWHM} > 1000$\kms\ are rarely seen in star-forming galaxies, even in the most powerful starbursts \citep[e.g.][]{amorin_complex_2012, amorin_ubiquitous_2024, crespo-gomez_rioja_2025, zamora_physical_2025, Marques-Chaves_extremely_2026}, but are more common in AGN-driven outflows \citep[e.g.][]{manzano-king_agn-driven_2019, cooper_high-velocity_2025, bertola_ga-nifs_2025}.
In that framework, \on\ lies in the most extreme ionised outflows driven by AGNs (See Fig. \ref{fig:vout}). 

The outflowing velocities observed in \on\ are significant, and stronger than most warm outflows \citep[e.g.][]{nandi_stratification_2026}. In order to estimate if this outflow could lead to the quenching of the galaxy \citep[as discussed in][]{taylor_jwst_2026}, we need to estimate the mass loading factor, i.e. the ratio between the outflowing mass rate and the SF rate. For that, we first need to estimate the outflowing ionised mass from the galaxy, for which we use the luminosity to mass relationship from \citet{colina_ic_1991}, with $n_e$ the electron density in \cmc\ and $L(\Oiiib)$ the \Oiiib\ luminosity in \ergs:
\begin{equation}
    M_g = \frac{1.21\times10^{-33} L(\Oiiib)}{n_e}
    \label{eq:outflow_mass}
.\end{equation}

Then, to estimate the mass rate, we use the standard equation \citep[e.g.][]{alvarez-marquez_detection_2021} introduced below, which relates the outflowing mass rate with the ionised mass, $M_g$ (in \msun), the maximum outflowing velocity, $v_{\rm out} = \abs{\Delta v} + 2v_{\rm FWHM}/2.355 = \abs{\Delta v} + 2\sigma_{\rm out}$ \citep[with $\Delta v$ the velocity difference between the two peaks and $\sigma_{\rm out}$ the width of the outflowing component, following][]{Genzel_sins_2011}, the size of the outflow, $d$ (in kiloparsecs), and a geometrical parameter, $K$, to quantify if the galaxy is elongated or circular. In our case, Fig.~\ref{fig:phot_cutout} shows an elongated object; therefore, we set $K=1$ \citep[see][]{lutz_molecular_2020, alvarez-marquez_detection_2021}. The equation is described as follows: 

\begin{equation}
    \dot{M}_g = K\frac{M_g v_{\rm out}}{d}
    \label{eq:mass_rate}
.\end{equation}

Because of the limited morphological information, we used rough estimates of $d = 0.7\arcsec = 2.46$kpc (de-lensed) , corresponding to the size of the outflow in the 2D spectrum.
We computed the mass loading factor as the ratio of the outflowing ionised mass rate and the ${\rm SFR}_{\ha}$ estimated from \citet{kennicutt_star_2012}:

\begin{equation}
    \eta = \frac{\dot{M}_g}{{\rm SFR}_{\ha}} = 10^{41.27 - \log L(\ha)} \dot{M}_g
    \label{eq:loading_factor}
.\end{equation}

\begin{table}[t]
    \centering
    \tiny
    \caption{Kinematic properties of the ionised gas in the host component.}

    \begin{tabularx}{\columnwidth}{lXll}
    \toprule
    Property & Units & Outflow 0 & Outflow 1\\
    \midrule
    $\sigma_{\rm out}$ & \kms & $325.63 \pm 13.89$ & $2271.36 \pm 241.30$ \\
    $\Delta v$ & \kms & $-58.48 \pm 5.16$ & $594.29 \pm 127.06$ \\
    $v_{\rm out}$ & \kms & $709.74 \pm 28.25$ & $5137.01 \pm 499.04$ \\
    $v_{\rm out}$ $^\dagger$ & \kms & $441.91 \pm 17.15$ & $3268.82 \pm 311.24$ \\
    $M_g$ & $10^{6} \msun$ & $6.91 \pm 6.43$ & $>2.70 \pm 2.53$ \\
    $\dot{M}_g$ & \msunyr & $2.04 \pm 1.90$ & $>5.78 \pm 5.43$ \\
    $\eta$ & --- & $0.05 \pm 0.04$ & $>0.13 \pm 0.12$ \\
    $\dot{E}_{\rm out}$ & $10^{41}$\ergs & $3.24 \pm 3.03$ & $>480.62 \pm 460.97$ \\
    \bottomrule
    \end{tabularx}
    
    \tablefoot{The properties are corrected for dust attenuation. $^\dagger$Alternative measurement of the maximum velocity from $v_{\rm out} = \abs{\Delta v} + v_{\rm FWHM}/2$ \citep[e.g.][]{Rupke_outflows_2005, arribas_ionized_2014}. The large uncertainties on $\dot{M}_g$ and subsequent results are driven by the large uncertainties on $n_e(S^+)$.}

    \label{tab:kinematics}
\end{table}

Using the dust-corrected line fluxes, we obtain mass loading factors of $\eta = 0.05 \pm 0.04$ for the first outflow and $\eta > 0.12$ for the large outflow. Since both values are well below unity, the mass ejected by the AGN is insufficient to counteract SF within the galaxy; therefore, these outflows are unable to quench this galaxy. Note that these mass loading factors are similar to those of the LRD in \citet{cooper_high-velocity_2025}, whose object had the fastest ionised outflow of their catalogue, with one of the lowest mass loading factors.  
Finally, we inferred the kinetic energy of the outflows using $\dot{E}_{\rm out} = \frac{1}{2}\dot{M}_g v_{\rm out}^2$ which gives $\dot{E}_{\rm out} = (3.24 \pm 3.03)\times10^{41}$\ergs\ and $\dot{E}_{\rm out} > 4.8\times10^{43}$\ergs. For such extreme outflowing velocities, the energy output is rather small. Comparatively, \citet{bertola_ga-nifs_2025} found $\dot{E}_{\rm out} \sim 5.2\times10^{44}$\ergs\ for an AGN with similar $v_{\rm out}$\footnote{Re-calculated using their equation and the $v_{\rm max}$ in their Table 2 instead of their definition of $v_{\rm out} = \max(v_{10},v_{90})$.}.
These results show that, despite the fast outflow observed in \oiii, the impact on the galaxy is rather small. 
A summary of the kinematic properties of the galaxy can be found in Table \ref{tab:kinematics}.

\begin{figure*}[t]
    \centering
    \includegraphics[width=0.33\linewidth]{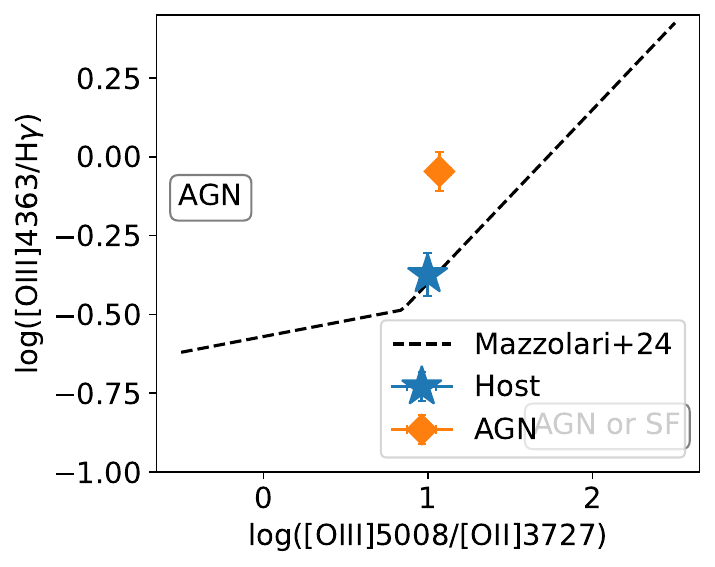}
    \includegraphics[width=0.33\linewidth]{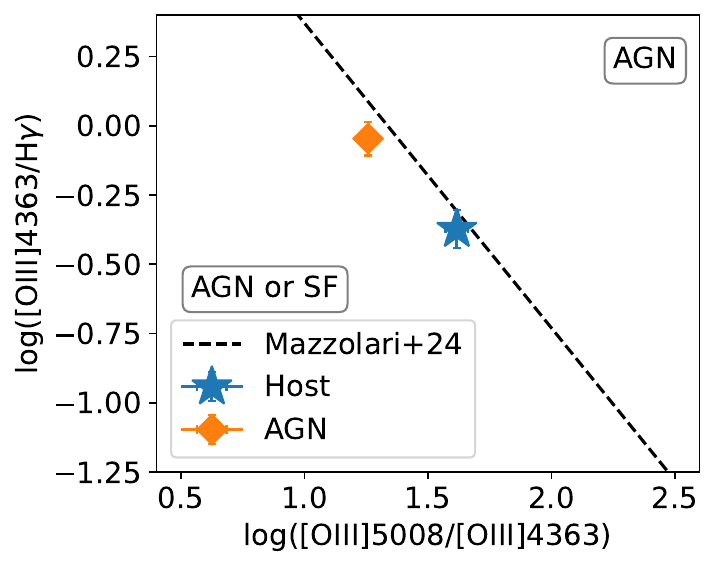}
    \includegraphics[width=0.33\linewidth]{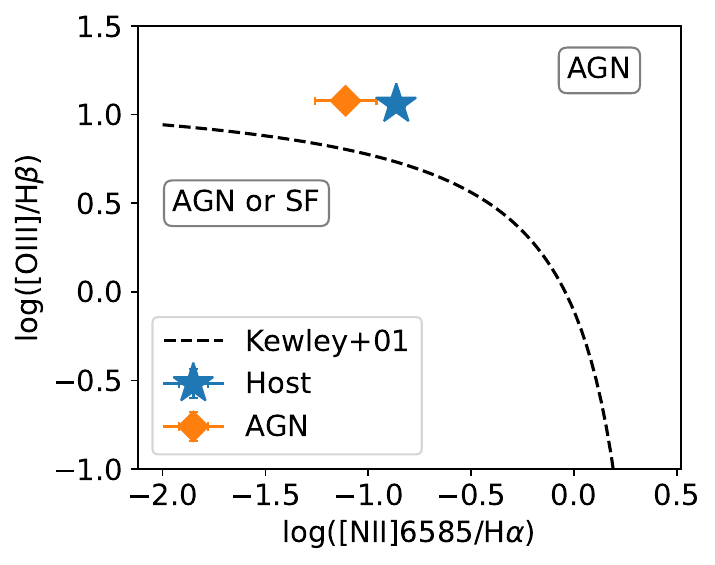}
    \caption{Line diagnostic figures for \on. Measurements are performed on the narrow lines of both the AGN and host components. Left and centre: Diagnostics based on \Oiiia\ from \citet{mazzolari_new_2024}. Right: BPT diagnostics from \citet{kewley_theoretical_2001} showing high compatibility with an AGN. \oiii\ corresponds to the sum of \Oiii.}
    \label{fig:lines_diagnostics}
\end{figure*}

We finally note the outflow is also visible in the AGN component. While slower and less significant, it is observed in \oiii\ and \ha\ as well. This is most probably due to contamination of the AGN spectrum with the host galaxy. We further note that the line profile from \Oiiib\ in the AGN is complex and cannot be explained by simple Gaussians. We shall now discuss the properties of the AGN.

\subsection{AGN properties}\label{sec:res_agn}

\subsubsection{Empirical evidence for an AGN}
We identified this component as an AGN based on multiple signatures. First, the \ha\ profile is compatible with a Type-1 AGN BLR. Then, as shown in Fig. \ref{fig:lines_diagnostics}, \on\ is clearly identified as an AGN in a BPT diagram \citep{baldwin_classification_1981, kewley_theoretical_2001} and more recent AGN line diagnostics \citep{mazzolari_new_2024}. Finally, we also note the massive outflows observed in the host, which are most probably produced by an AGN (see Fig.~\ref{fig:vout}). We therefore assume that we are observing an AGN, and in the following subsections, we discuss its observed characteristics. Furthermore, we show that \on\ also strongly resembles LRDs. The main inferred properties are listed in Table \ref{tab:general_properties}.

\subsubsection{Black hole mass}\label{sec:bhm}
In this subsection, we first make the assumption that the observed exponential wings in \ha\ and \hb\ are solely due to gas motion because of the black hole. However, we then discuss that the estimations used here most probably do not translate well to \on. Nonetheless, they provide interesting anchor points for comparison with other objects.

We estimated the black hole mass \BHM\ following \citet{reines_relations_2015}, which uses the single-epoch virial mass estimator from \citet{reines_dwarf_2013} and an estimated scale factor of $\epsilon=1.075$. This estimator uses the FWHM from the Gaussian fit of \ha. However, as seen earlier, our \ha\ profile is better fitted by an exponential profile, which has narrower FWHM, leading to an underestimation of the mass. Nonetheless, we obtain a de-lensed black hole mass $\log(\BHM/\msun) = 6.96 \pm 0.50$. 
Due to the lack of photometric constraints, we cannot directly measure the stellar mass $\mstar$ of \on. However, using the average mass-metallicity calibration at $z=6-8$ from \citet{chemerynska_extreme_2024} and the metallicity of the host determined above, we estimate a de-lensed host stellar mass of $\log(\mstar/\msun) = 8.21 \pm 0.23$. This stellar mass is low for a system hosting such a massive black hole. For example, applying the local scaling relations from \citet{reines_relations_2015} would predict a black hole mass 2–3 orders of magnitude lower than measured. However, \on\ aligns well with the behaviour of the  LRD population \citep{jones_relationship_2025}, which typically shows a higher black hole mass to stellar mass ratio, compared to more standard AGNs.

The black hole mass estimates rely on the assumption that the broadening of \ha\ is driven exclusively by gas motion. However, this assumption may not apply to LRDs. While some studies argue that the observed exponential line profiles in LRDs could still arise from gas motion in stratified BLR clouds \citep[e.g.][]{scholtz_little_2026}, others attribute these profiles to electron and/or resonant scattering due to extreme densities in the BLR, which would also produce an exponential shape \citep[e.g.][]{Kokorev_deepest_2026, rusakov_little_2026, chang_impact_2026}. In the latter scenario, the black hole mass estimates would be invalid, as the line broadening would not reflect gas kinematics \citep[e.g.][]{Naidu_black_2026, degraaff_little_2026}.

\subsubsection{Electron temperature, electron density, and abundances of AGN component and obscuration}
In order to characterise the gas surrounding the AGN, we applied a similar analysis to that used for the host component. We used \Oiiib\ and \Oiiia\ to measure $T_e(O^{+2})$, and \Sii\ for $n_e(S^+)$. We did not expect to observe \sii\ from the BLR, as the densities are higher than their critical density. Therefore, we only considered a narrow component for this doublet \citep[e.g.][]{kewley_theoretical_2019}. We used \texttt{PyNeb} to compute these quantities, and fitted them to obtain a coherent solution. We measured a slightly higher electron temperature, $T_e(O^{+2}) = 24357_{-940}^{+484}$K, than for the host, but similar electron densities: $\log(n_e(S^+)/{\rm cm}^{-3}) = 3.35_{-0.56}^{+0.50}$.

Finally, we measured the Balmer decrement in the AGN. While the narrow lines are compatible with the absence of dust, the BLR gives $\ha/\hb = 16.35 \pm 2.55$. This value is extreme compared to lower-redshift AGNs and bright quasi-stellar objects (QSOs) \citep[e.g.][]{sun_little_2026}, but typically observed in LRDs \citep[e.g.][]{Torralba_warm_2026, degraaff_little_2026}. If caused by large quantities of dust, it would correspond to an extreme attenuation \citep[e.g.][]{kocevski_hidden_2023}. But the absence of cold and warm dust re-emission in the MIR/FIR flux of LRDs rules out large quantities of dust \citep[e.g.][]{casey_upper_2025, setton_confirmed_2025}, favouring alternative scenarios such as a compact dust reservoir or resonant and electron scattering effects in dust-poor dense gas envelopes \citep[e.g.][]{casey_upper_2025, Naidu_black_2026, degraaff_little_2026, chang_impact_2026, rusakov_little_2026}.

\subsubsection{A forest of Fe II lines}\label{sec:feii}

One of the features of our AGN component is the plethora of rest-frame optical \feii\ lines detected. Figures \ref{fig:agn_spec_fig} and \ref{fig:comp_lrd} show the fits for these emission lines, and their fluxes are listed in Table \ref{tab:agn_iron_forest}. While weak, several of these lines are detected at $>3 \sigma$: \Feii{4245}, \Feii{4287}, \Feii{4414}, and \Feii{5159}. 
So far, this \feii\ forest has only been observed in a few LRDs, such as in local analogues \citep{ji_lord_2026, lin_discovery_2026} and at high redshifts \citep{lambrides_discovery_2025, deugenio_irony_2025, Torralba_warm_2026}. These lines are typically very weak, requiring high-resolution and deep observations to be properly detected.

We further note that the \Feii{5159} line is blended with the higher-level \Fevii{5160}\ line. This degeneracy between the two blended lines was observed and discussed in previous AGN or LRD studies \citep[e.g.][]{deugenio_irony_2025, lambrides_discovery_2025, Wang_missing_2026}. For the same reason as in ‘Irony’ \citep{deugenio_irony_2025}, we believe this emission line to be \feii\ rather than \fevii: we observe the same \feii\ forest, and we do not detect \Heii{4686} in the AGN component. Since the ionisation potential of He$^+$ (54 eV) is lower than that of Fe$^{+6}$ (99 eV), we would expect to see \Heii{4686} if \Fevii{5160}\ was present, which is not the case.

\subsubsection{Inflowing gas}

\begin{figure}
    \centering
    \includegraphics[width=0.8\linewidth]{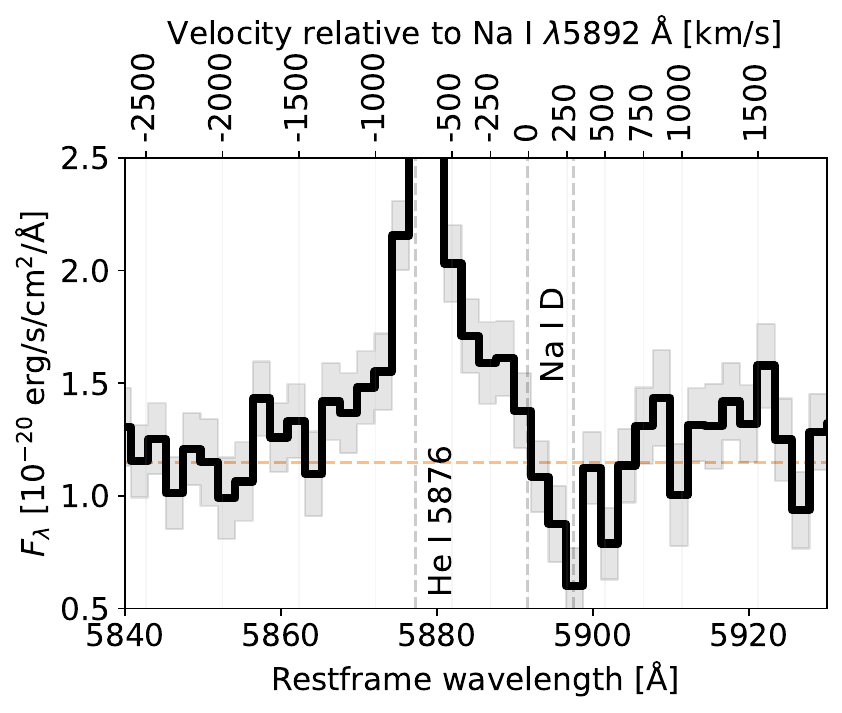}
    \caption{Na {\sc i d} absorption feature in the AGN component with top axis centred around the \Nait line.}
    \label{fig:naid}
\end{figure}

In addition to emission lines, the AGN spectrum exhibits an absorption feature at the expected position of \nai\ (Fig.~\ref{fig:naid}), which has also been reported in a few LRDs. Although not as deep as in some previously observed systems \citep[e.g.][]{deugenio_irony_2025, ji_lord_2026, lin_discovery_2026}, it is clearly detected in our spectrum. Characterising this feature is, however, challenging, as the \nai\ profile is likely contaminated by the broad emission from \HeiNA. 
In objects such as \citet{ji_lord_2026} (see Fig.~\ref{fig:comp_lrd}c), where kinematic signatures are less pronounced, \HeiNA\ and \nai\ are well separated. In our case, the absorption appears redshifted by $\simeq 280$\,\kms, which is indicative of inflowing gas towards the AGN along the line of sight. This inflow is also consistent with the inability of the fast outflow to quench the host galaxy, as in addition to the low mass loading factor, a gas reservoir flows towards it.

\subsubsection{Balmer break and black-body}\label{sec:bb_and_blackbody}

\begin{figure}[t]
    \centering
    \includegraphics[width=\linewidth]{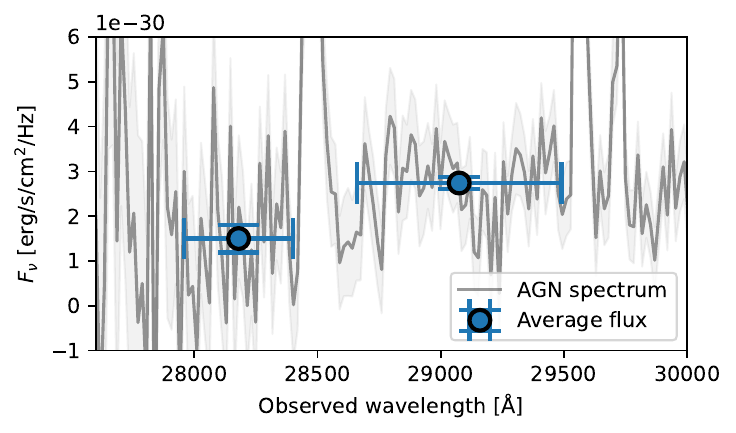}
    \caption{Balmer break of the AGN component in the nod difference reduced spectrum.}
    \label{fig:balmer_break}
\end{figure}

Due to the limited spectral coverage, we cannot probe the Balmer break with traditional estimators \citep[e.g. 3751-4198Å vs 3145-3563Å from ][]{Curtis-Lake_spectroscopic_2023}. In addition, as described in the method, the spectrum based on the global noise reduction has issues on the edges of the spectrum. Therefore, we probed the region redwards of the Balmer break using the nod difference subtraction spectrum of the AGN, which does not suffer from the same sky background issue (see Sect.~\ref{sec:data} and App.~\ref{app:nod_vs_gsky}). In order to avoid edge noise and emission lines, we defined the bluer region as $\num{27960}-\num{28400}$Å (observed, rest-frame $\num{3660}-\num{3717}$Å), and the redder region as $\num{28660}-\num{29490}$Å (observed, rest-frame $\num{3751}-\num{3860}$Å). We then measured the average flux in these two regions and obtained a ratio of $1.82 \pm 0.39$, shown in Fig. \ref{fig:balmer_break}. This result is significant, and while it does not probe the Balmer Break, it shows a significant decrease redwards of the Balmer break, as is observed in other LRDs \citep[e.g.][]{Naidu_black_2026, degraaff_little_2026}.

\begin{figure}[t]
    \centering
    \includegraphics[width=\linewidth]{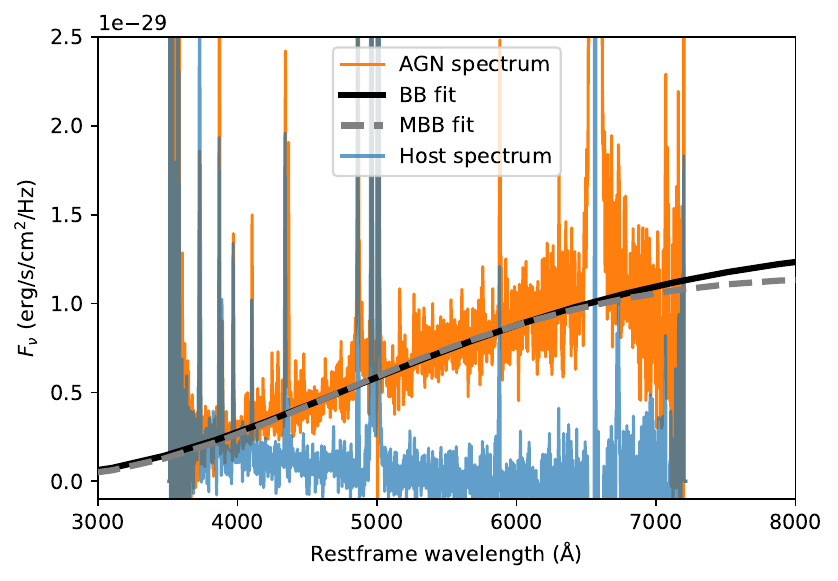}
    \caption{Fit of the AGN component with a simple black-body \citep[BB, following][]{Naidu_black_2026, sun_little_2026} and a modified black-body \citep[MBB, following][]{degraaff_little_2026}. We also report the host component as a comparison.}
    \label{fig:black_body}
\end{figure}

In addition, previous studies also observed that black-body models fit LRDs well \citep[e.g.][]{Naidu_black_2026, degraaff_little_2026, sun_little_2026}, calling this model the black hole star \citep[\bhs,][]{Naidu_black_2026}. To test this, we defined two simple models, the standard black-body \citep[BB, e.g.][]{Naidu_black_2026, sun_little_2026} and the modified black-body \citep[MBB, following][]{degraaff_little_2026}, which includes an additional component to account for the contribution of the galaxy. These models are defined by three parameters: the temperature, $T$, the scaling factor, $A$, and the slope, $\beta$, which was set to $\beta = 0$ for the BB model. Following \citet{degraaff_little_2026}, we defined our models as

\begin{equation}
    f_\nu = A\times B_\nu(T)\times\left(\frac{5500{\rm Å}}{\lambda}\right)^\beta 
    \label{eq:black_body}
,\end{equation}

\noindent where $\lambda$ is the rest-frame wavelength in Å and $B_\nu(T)$ is the general definition of the black-body spectrum. We fitted these two models to our AGN component (nod difference spectrum with emission lines masked) and show it in Fig. \ref{fig:black_body}. We first note that, while the BB model is well parametrised, the MBB model does suffer from degeneracies between the scale, $A$, and slope, $\beta$. Nonetheless, the reduced $\chi^2$ and the information criteria are similar between the two parametrisations. The models have a temperature of between $T\sim4000-5000$K, which is comparable to the previous observations \citep{degraaff_little_2026, sun_little_2026}. The resulting bolometric luminosities of $L_{\rm blackbody} \sim 10^{45}$\ergs\ are on the higher end of previously observed LRDs \citep{degraaff_little_2026}, but are compatible with their results. This difference might be explained by the lack of spectral coverage redwards of \ha. Our AGN component therefore matches well with the LRD population. The fitted parameters can be found in Table \ref{tab:blackbody_fit}.

\begin{table}[]
    \centering
    \tiny
    \caption{Black-body fitting parameters for the AGN component}.

    \begin{tabularx}{\columnwidth}{XXll}
    \toprule
    Parameters & Units & Black-body & Modified Black-body \\
    \midrule
    $T$ & K & $5064.73\pm46.08$ & $4152.19\pm349.45$ \\
    $A$ & $10^{25}$ & $2.28\pm0.11$ & $7.20\pm3.85$ \\
    $\beta$ & --- & $0^\dagger$ & $1.17\pm0.55$ \\
    $\chi^2_\nu$ & --- & 1.09 & 1.09 \\
    BIC & --- & 85.80 & 87.66 \\
    AIC & --- & 76.38 & 73.53 \\
    $L_{\rm blackbody}$ & $10^{45}$ \ergs & $1.37 \pm 0.05$ & $1.12 \pm 0.91$ \\
    \bottomrule
    \end{tabularx}

\tablefoot{$T$ is the black-body temperature, $A$ is the de-lensed scaling factor (including the conversion for steradians), $\beta$ is the slope modifier for the modified black-body, $\chi^2_\nu$ is the reduced $\chi^2$, and BIC and AIC are the Bayesian and Akaike information criteria. $L_{\rm blackbody}$ is the de-lensed bolometric luminosity of the models. $^\dagger$ $\beta$ is fixed to 0 for the standard black-body.}

    \label{tab:blackbody_fit}
\end{table}

\section{Discussion}

\subsection{Similarities with little red dots}\label{sec:disc_lrd}

\begin{figure*}
    \centering
    \includegraphics[width=\linewidth]{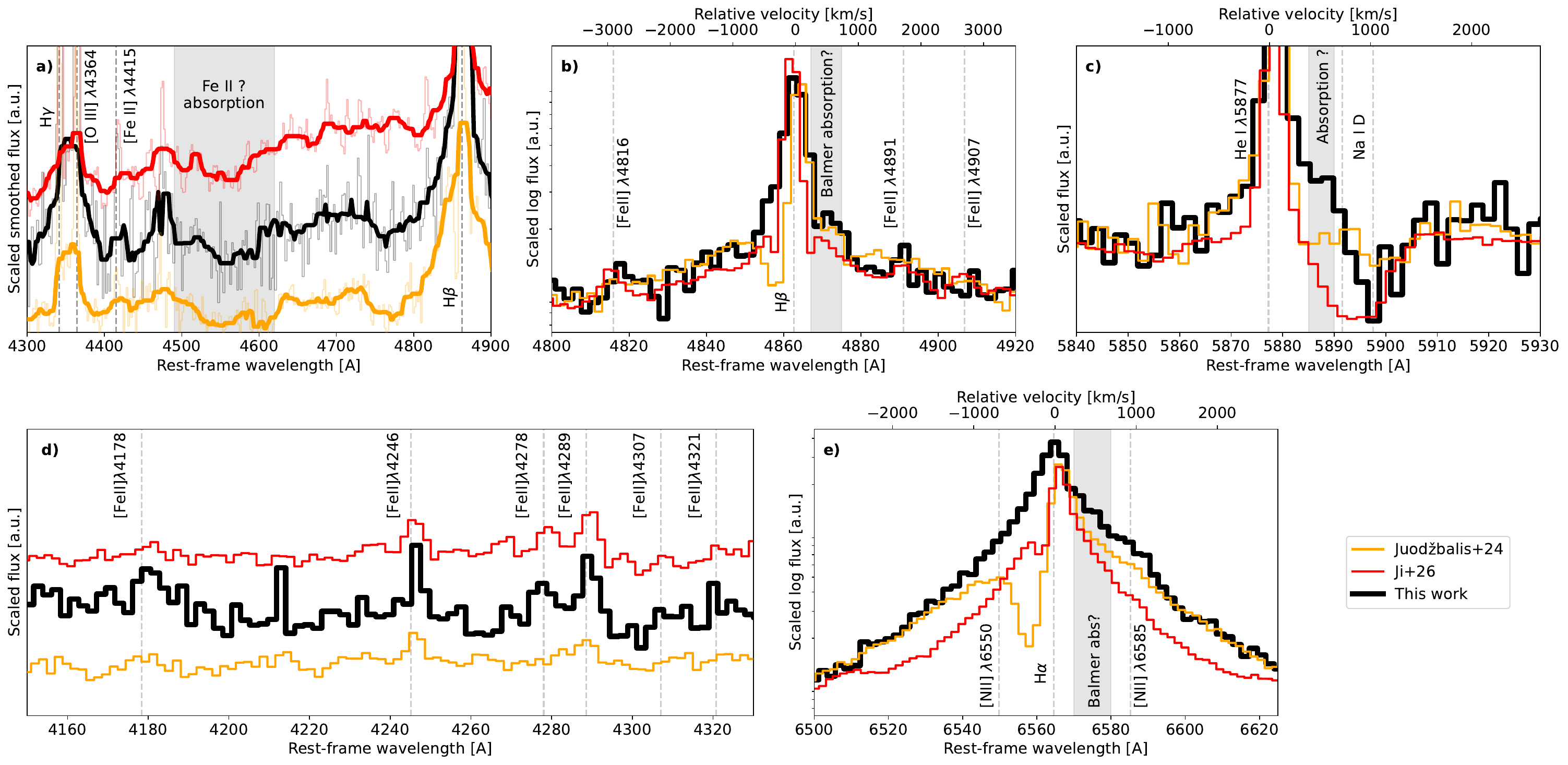}
    \caption{Comparison between the LRD-like features in \on, and two deep and comparable LRDs \citep{juodvbalis_jades_2024, ji_lord_2026}. The fluxes are scaled: a) shows the \recombination{Fe}{ii} absorption forest, b) \hb\ in log-scale, c) \Hei{5875} and the \nai\ {\sc d} complex, d) the blue part of the \feii\ forest, and e) \ha\ in log-scale.}
    \label{fig:comp_lrd}
\end{figure*}

Little red dots likely constitute a diverse population of compact extragalactic sources that do not all exhibit the same spectral properties. Nevertheless, several common characteristics appear to emerge: a V-shaped continuum, prominent broad wings in the Balmer lines, a strong absorption feature around $\sim4550$\,\AA, and the presence of \feii\ emission \citep[e.g.][]{matthee_little_2024, furtak_high_2024, greene_uncover_2024, juodvbalis_jades_2024, deugenio_irony_2025, Kokorev_deepest_2026, perez-gonzalez_little_2026, ji_lord_2026}.
\on\ shows multiple features in common with typical LRDs. We shall now discuss these different features. To illustrate it, we compare the spectrum of \on\ with those of two well-studied LRDs in Fig.~\ref{fig:comp_lrd}, SDSS J102530.29+140207.3, also named ‘The Egg’ in \citet{lin_discovery_2026} or ‘Lord of LRDs’ in \citet{ji_lord_2026} (we keep the name of the latter for clarity), a low-redshift ($z\sim0.1$) LRD, and the AGN Rosetta Stone from \citet{juodvbalis_jades_2024} at $z\sim2.26$. The three spectra have comparable signal-to-noise ratios and spectral resolutions.

One of the initial criteria for LRD was the overall shape of the SED: a V-shaped SED \citep[e.g.][]{matthee_little_2024}. By  fitting our AGN component (nod difference) with $f_\lambda\sim\lambda^{\beta_{\rm opt}}$, we obtain a positive slope of $\beta_{\rm opt} = 0.37 \pm 0.05$. Using F125W and F160W from the HST photometry (excluding F105W, which might include Lyman-$\alpha$), we can approximate the UV slope from F125W and F160W HST filters: $\beta_{\rm UV} = -1.60 \pm 0.04$. These slopes match well the V-shaped description and set our galaxy in the category of LRDs \citep[e.g.][]{brazzini_little_2026}. In addition, recent papers highlighted the black-body shape of LRDs \citep[e.g.][]{Naidu_black_2026, degraaff_little_2026, sun_little_2026}, which is well reproduced in our AGN component (see Sect.~\ref{sec:bb_and_blackbody}), and our temperatures of $T=4000-5000$K compare well to other LRDs \citep[e.g.][]{degraaff_little_2026}. Our observations of a Balmer-break also reinforce the comparison. Furthermore, the limited Spitzer data points enable us to tentatively exclude a red AGN \citep[e.g.][]{fei_direct_2026} due to the non-detection of the IRAC3 and 4 bands, as these objects are expected to shine in the rest-frame infrared, and not drop as LRDs do \citep[e.g.][]{degraaff_little_2026}. Therefore, the general morphology of our object fits the typical LRD well.

Another striking feature is the presence of the absorption feature around $\sim4550$ Å (see panel a) in Fig.~\ref{fig:comp_lrd}), which appears in many individual LRDs and in deep stacks of LRDs \citep{perez-gonzalez_little_2026} and individual spectra \citep[e.g.][]{ji_lord_2026, deugenio_irony_2025}.
A plausible explanation would be metallic absorption from permitted \recombination{Fe}{ii} lines \citep{perez-gonzalez_little_2026}, as many of these lines are identified around $\sim4550$Å in quasars \citep[e.g.][]{Boroson_emission-line_1992, Veron-Cetty_unusual_2004, Dong_broad-line_2008, kovacevic_analysis_2010, kovacevic-dojcinovic_searching_2025}. 

We also report the presence of numerous \feii\ emission lines, which are typically observed in LRD spectra of sufficient depth and spectral resolution 
\citep[e.g.][]{juodvbalis_jades_2024, deugenio_irony_2025, lambrides_discovery_2025, ji_lord_2026, lin_discovery_2026}. These requirements make it difficult to observe these lines at higher redshift, explaining the low number of detections in these galaxies \citep{lambrides_discovery_2025, deugenio_irony_2025}. \on\ is one of these few high-redshift ($z \sim 6.64$) objects with significant detections of multiple \feii\ lines, as discussed earlier (Sect.~\ref{sec:feii}). 

Another feature is the presence of metallic absorption in deep spectra of LRDs, such as \nai, \mgi, \caii\ ,  or \ki\ \citep[e.g.][]{juodvbalis_jades_2024, deugenio_irony_2025, ji_lord_2026, lin_discovery_2026}. In \on, we only observe a \nai\ absorption, as seen on panel c) of Fig.~\ref{fig:comp_lrd}, but this gives another hint of an LRD.

Finally, we also observe exponential wings in the \ha\ emission line (see Sect. \ref{sec:fit_agn}), as shown by the triangular-shaped \ha\ profile in panel e) of Fig. \ref{fig:comp_lrd}. This is comparable to some LRDs \citep[e.g.][]{rusakov_little_2026}, including the Rosetta Stone and the Lord of LRDs. 
In \hb, this exponential profile is not as strong due to the strong Balmer decrement and the stronger contribution of the narrow \hb\ component to the emergent line profile.
Furthermore, we note the absence of clear Balmer absorption in the BLR of \ha\ and \hb\ that are clearly shown in the Rosetta Stone and Lord of LRDs. 
These absorption features are only expected in extreme densities, but depend on the complex conditions of the gas and the contribution of the narrow emissions from the underlying component; therefore, many configurations of the absorption are observed \citep[e.g.][]{juodvbalis_jades_2024, ji_blackthunder_2025, deugenio_irony_2025, lambrides_discovery_2025, ji_lord_2026, lin_discovery_2026, rusakov_little_2026}, including no visible absorption \citep[e.g.][]{rusakov_little_2026, lin_discovery_2026}. This absence of absorption can easily be explained in the situation where the absorption is at the systemic redshift. In that case, it would be compensated by the narrow component, which might be the case in our situation. Alternatively, it has been shown by \citet{deugenio_jades_2026} that medium-resolution gratings such as G395M can hide such features, while being observed at higher resolution, with G395H for example. Note that such absorption at high densities was also observed in \hei\ \citep[e.g.][]{Kokorev_deepest_2026}. However, we note that, while we do not observe clear absorption features in \ha, \hb, and \hei, we observe some surprising common behaviour. In \hb\ and \singlet{\hei}{5877}, we observe a sharp drop of the profile at $\sim500$ \kms, with a redward small absorption, possibly indicating a common origin. This exact feature is not visible in \ha, but we note that the red wing of \ha\ does also have a sharp drop around the same relative velocity. In the end, all these possible absorptions are very tentative. They require higher-resolution observations to be properly analysed, and do not represent conclusive arguments in favour of an LRD.

\subsection{Nature of the ionised outflow in \on}\label{sec:nature_outflow_lrd}

The frequency of ionised outflows in LRDs remains poorly constrained, as most have been observed with low-resolution spectroscopy (e.g. PRISM) or with a higher resolution but shallower spectroscopy \citep[e.g.][]{degraaff_rubies_2025}. 
So far, ionised outflows have only rarely been detected, in LRDs observed with both higher resolution and sufficient signal-to-noise \citep{juodvbalis_jades_2024, cooper_high-velocity_2025, chen_z_2026, deugenio_jades_2026}. While theoretical models predict that outflows may be trapped within the dense gas envelope surrounding LRDs and re-emitted as black-body radiation \citep{kido_black_2025}, their detection in some systems may suggest that the envelope is at least partially clumpy or has been disrupted, allowing a fraction of the gas to escape. 
Interestingly, the extreme outflow of \on\ is contrasted by its minor impact on the galaxy.
Indeed, measurements of the mass loading factor and kinetic energy from the outflowing and inflowing gas show a minimal impact, much smaller than other outflows with comparable velocities \citep[e.g.][]{bertola_ga-nifs_2025}. This reinforces the argument for a disrupted envelope, which would only release a tiny fraction of the gas.

\begin{figure}[t]
    \centering
    \includegraphics[width=\linewidth]{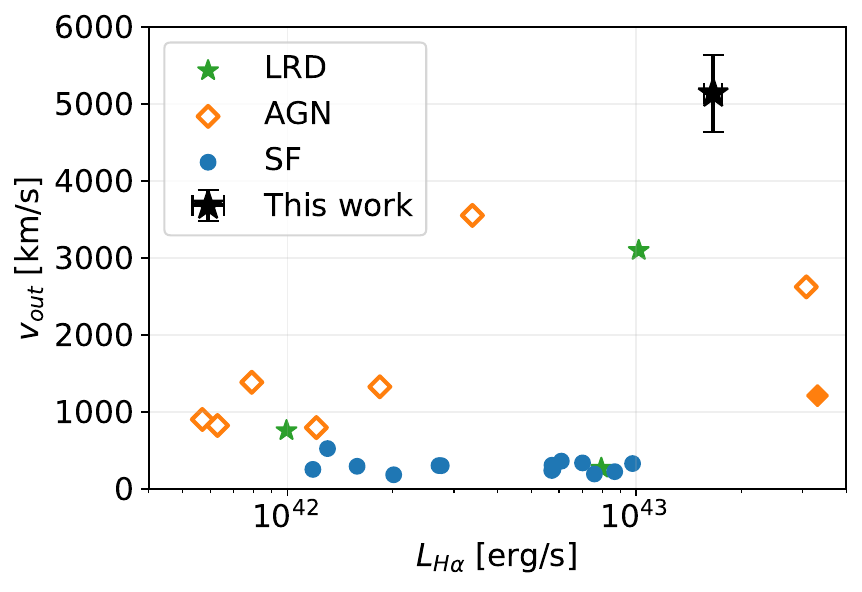}
    \caption{Comparison between the outflowing velocities and total \ha\ flux from SF feedback, AGN feedback, and for the few detected outflowing LRDs. The full markers show the total \ha, while empty markers correspond to the broad components only. All the velocities were harmonised to $v_{\rm out} = \abs{\Delta v} + 2\sigma_{\rm out}$. All the values are reported in Tab.~\ref{tab:vout}.}
    \label{fig:vout}
\end{figure}

Figure \ref{fig:vout} shows the maximum outflowing velocity, $v_{\rm out}$, against the total \ha\ luminosity of \on,
compared to observations of star-forming galaxies, AGN, and LRDs.
In this context, \on\ stands out as a clear outlier due to its extreme outflow velocity. Such high-velocity outflows are observed almost exclusively in AGN-dominated systems and are significantly stronger than those typically found in star-forming galaxies, including even the most extreme starbursts \citep[e.g.,][]{alvarez-marquez_detection_2021, rodriguez_del_pino_ga-nifs_2026}. The LRDs presented by \citet{chen_z_2026} and \citet{cooper_high-velocity_2025} also lie within the AGN-dominated region of the diagram, while the LRD from \citet{deugenio_jades_2026} lies at lower velocity. Although the current sample of LRDs with detected outflows remains limited, their location in this parameter space strongly supports an AGN-driven origin.

Interestingly, we also observe a discrepancy of $\abs{\Delta v_{\rm sys}} \sim 80$\kms\ between the systemic redshift of the host and the AGN. Due to the limited spatial information, we cannot exclude that \on\ is a merging system, as the velocity difference and spatial separations are in line with observations and simulations of mergers \citep[e.g.][]{Patton_new_2000, Ventou_new_2019}. We do not detect any additional companion for the LRD, but it might simply be too faint \citep[e.g. about $\sim43-80\%$ of LRDs have at least one UV-bright companion][]{Baggen_connecting_2026}. Furthermore, we can also suggest the possibility of the presence of multiple black holes; for example, one in the host causing the outflow, and one in the LRD. However, due to the limited spatial data available, an accurate assessment of these hypotheses requires further observations.

In conclusion, we found strong evidence that the observed fast ionised outflow in \on\ is driven by AGN activity. The similarities between this object and the diverse class of LRDs, including spectral features, emission line profiles, and continuum shape, suggest that this LRD hosts an active AGN. To date, most LRD studies have relied on low-resolution PRISM data, which may limit the detection of outflows. Future medium- or high-resolution spectroscopic observations of LRDs will be important to determine whether such extreme outflows are a common feature of these systems.

\section{Summary and conclusion}

In this work, we present \on, a $z = 6.64$ galaxy behind the Abell S1063 cluster, observed as part of the deep JWST/NIRSpec GLIMPSED survey. This galaxy stands out due to its spatially resolved structure showing a compact AGN component exhibiting classic LRD features, and an extended host galaxy exhibiting a very fast ionised outflow driven by the AGN. Using medium-resolution spectroscopy, we separated the spectra of the AGN and host, revealing a wealth of emission and absorption features.
The main results from their analysis are the following:

\begin{itemize}
    \item The spectrum of the host galaxy exhibits an extreme ionised outflow, mainly observed in \Oiii, \ha\ , and \hb. This outflow is modelled with two components with maximal outflowing velocities of $v_{\rm out} \sim 709$\kms\ and $v_{\rm out} \sim 5137$\kms\ , respectively. The first outflow is observed at approximately the same redshift as the narrow component, but the second outflow is blueshifted by $\Delta v \sim 594$\kms. The narrow component is compatible with no dust attenuation, while the two outflowing components are compatible with some attenuation, hinting at expulsion of dust with the outflow. 
    The slower outflowing component does not significantly affect the galaxy, as the mass loading factor only reaches $\eta \sim 5\%$. However, the faster outflowing one lacks \hb\ detection that only provides a lower limit on the dust attenuation, and hence a lower limit of $\eta > 10\%$. Unless the outflow is significantly dusty, we do not expect a massive impact of the outflow on the host galaxy.
    The gas in the host component shows a relatively low oxygen abundance of $12+\log(O/H) \sim 7.96$ and slight nitrogen enhancements $\log(N/O)\sim-0.68$. Using a mass-metallicity relationship \citep{chemerynska_extreme_2024}, we estimate a de-lensed stellar mass of $\log(\mstar/\msun)\sim8.21$.
    \item The AGN component shows many typical features of LRDs. First, we observe the V-shaped SED. Then, we observe large exponential wings in \ha, which are significantly weaker in \hb. The Balmer decrement is too large to be explained by dust attenuation, and is most probably caused by electron scattering instead. We infer a significant over-massive black hole mass, whose ratio with the stellar mass of the host places it 2-3 orders of magnitude higher than typical relationship \citep{reines_relations_2015}, well in line with typical LRD population \citep{jones_relationship_2025}. \on\ further displays multiple significant detections of forbidden \feii\ lines, which are only observed in LRDs with reasonable exposure time and resolution. The rest-frame optical continuum is also well fitted by a black-body, as earlier observed in other LRDs \citep{degraaff_little_2026}. The significant absorption observed around $\sim4550$ \AA\ rest-frame in the AGN spectrum is also a common feature of LRDs. Together, these features provide strong evidence that \on\ hosts an LRD.    
\end{itemize}

Only a few LRDs have detected outflows, likely due to the low resolution or shallow depth of most spectra. In these cases, the extreme velocities are only consistent with AGN-driven outflows, providing compelling evidence that LRDs host AGNs. However, the current statistical sample of outflowing LRDs remains too small to generalise to the broader LRD population. With its extreme outflow and detailed spectral features, \on\ offers a unique opportunity to probe LRDs in a regime that has, until now, been largely unexplored.

\begin{acknowledgements}
DK would like to thank Ilona Morel for the help with PyNeb, Elena Bertola and the anonymous referee for the improvements to the draft.
HA acknowledges support from CNES, focused on the JWST mission, and the Programme National Cosmology and Galaxies (PNCG) of CNRS/INSU with INP and IN2P3, co-funded by CEA and CNES and from the French National Research Agency (ANR) under grant ANR-21-CE31-0838.
This work is based [in part] on observations made with the NASA/ESA/CSA James Webb Space Telescope. The data were obtained from the Mikulski Archive for Space Telescopes at the Space Telescope Science Institute, which is operated by the Association of Universities for Research in Astronomy, Inc., under NASA contract NAS 5-03127 for JWST. These observations are associated with program \#9223. All the JWST data used in this paper can be found in MAST: \href{https://doi.org/10.17909/3me9-n627}{doi:10.17909/3me9-n627}.
\textit{Softwares}: LMFit \citep{newville_lmfit_2025}, Astropy \citep{collaboration_astropy_2022}, Numpy \citep{harris_array_2020}, Matplotlib \citep{hunter_matplotlib_2007}, Scipy \citep{virtanen_scipy_2020}, msaexp \citep{brammer_msaexp_2023}.
\end{acknowledgements}

\bibliographystyle{aa}
\bibliography{references.bib}

\appendix

\section{Tables of emission lines}\label{app:table_emission_lines}

The measurement of emission requires careful treatment, due to the effects of the instrument. The width of a line is determined by the convolution of the intrinsic width and the instrumental width given by the resolving power $R(\lambda)=\lambda/\Delta\lambda=c/\Delta v$ of the instrument, where $\lambda$ is the wavelength, $\Delta\lambda$ is the minimum distinguishable wavelength, $c$ is the speed of light and $\Delta v$ is the minimum emission line velocity. Therefore, when modelling the emission line with a Gaussian, one needs to account for the instrument. The measured standard deviation $\sigma_{\rm tot}$ is therefore obtained by the following equation, where $\sigma_{\rm int}$ is the intrinsic width and $\sigma_{\rm instr}$ is the instrumental width, which varies with $\lambda$:

\begin{equation}
    \sigma_{\rm tot} = \sqrt{\sigma_{\rm int}^2 + \sigma_{\rm instr}^2} 
    = \frac{\lambda_{\rm line}}{2.355}\sqrt{\left(\frac{v_{\rm FWHM}}{c}\right)^2 + \left(\frac{1}{R(\lambda_{\rm line})}\right)^2}
    \label{eq:emission_line_total_width}
\end{equation}

In the second part of the equation, we replaced the Gaussians standard deviation with the parameters we report below. In addition, we define $\lambda_{\rm line}$ as the observed wavelength of the emission line.

We now report the tables of emission lines measured in \on. In the following table, the first column provides the name of the emission (or absorption) line of interest with its associated rest-frame wavelength in \AA. The second column reports the observed FWHM velocity dispersion $v_{\rm FWHM}$ for the lines and the final column reports the observed (not de-lensed) integrated flux of the line. Table \ref{tab:host_emission_lines} reports the measurement of the lines in the host component, and Table \ref{tab:agn_emission_lines} reports the emission (and absorption) lines from the AGN, except for the forbidden \feii\ lines which are reported separately in Table \ref{tab:agn_iron_forest}.

\begin{table}[ht]
\centering
\tiny
\caption{host emission line fits.}

\begin{tabularx}{\columnwidth}{Xll}
\toprule
Line & $v_{\rm FWHM}$ & Flux\\
(1) & (2) & (3)\\
 & km/s & $10^{-18}$ cgs\\
\midrule
\multicolumn{3}{c}{Narrow lines ($z=6.63785\pm0.00004$)} \\
\singlet{\oii}{3727.09} & $477.20\pm4.13$ & $4.50\pm0.47$ \\
\singlet{\oii}{3729.88} & $476.93\pm4.13$ & $2.97\pm0.44$ \\
\singlet{\hg}{4341.68} & $426.41\pm4.62$ & $2.55\pm0.31$ \\
\singlet{\oiii}{4364.44} & $424.84\pm4.64$ & $1.08\pm0.11$ \\
\Heii{4687.02} & $404.43\pm4.87$ & $0.29\pm0.09$ \\
\singlet{\hb}{4862.68} & $394.63\pm4.99$ & $5.21\pm0.34$ \\
\Hei{4923.304} & $391.44\pm5.03$ & $<0.31$ \\
\singlet{\oiii}{4960.30} & $389.53\pm5.06$ & $14.95\pm0.36$ \\
\singlet{\oiii}{5008.24} & $387.11\pm5.09$ & $44.70\pm1.07$ \\
\singlet{\nii}{6549.86} & $331.15\pm5.95$ & $0.74\pm0.06$ \\
\singlet{\ha}{6564.61} & $330.77\pm5.95$ & $16.11\pm0.73$ \\
\singlet{\nii}{6585.27} & $330.23\pm5.96$ & $2.21\pm0.18$ \\
\Hei{6679.995} & $327.84\pm6.01$ & $0.59\pm0.15$ \\
\singlet{\sii}{6718.29} & $326.89\pm6.02$ & $0.80\pm0.15$ \\
\singlet{\sii}{6732.67} & $326.54\pm6.03$ & $0.97\pm0.16$ \\
&&\\
\multicolumn{3}{c}{Outflow 0 ($z=6.63934\pm0.00022$)} \\
\singlet{\hg}{4341.68} & $787.43\pm31.85$ & $1.77\pm0.42$ \\
\singlet{\hb}{4862.68} & $770.69\pm32.54$ & $3.37\pm0.53$ \\
\singlet{\oiii}{4960.30} & $768.09\pm32.65$ & $4.40\pm0.63$ \\
\singlet{\oiii}{5008.24} & $766.86\pm32.71$ & $13.21\pm1.89$ \\
\singlet{\ha}{6564.61} & $740.02\pm33.89$ & $12.08\pm1.42$ \\
&&\\
\multicolumn{3}{c}{Outflow 1 ($z=6.62271\pm0.00562$)} \\
\singlet{\hb}{4862.68} & $5349.60\pm568.19$ & $<1.30$ \\
\singlet{\oiii}{4960.30} & $5349.23\pm568.23$ & $1.58\pm0.25$ \\
\singlet{\oiii}{5008.24} & $5349.05\pm568.25$ & $4.76\pm0.76$ \\
\singlet{\ha}{6564.61} & $5345.27\pm568.65$ & $4.78\pm1.02$ \\
\bottomrule
\end{tabularx}

\tablefoot{(1) Emission line name (2) Observed FWHM velocity. (3) Observed line flux in cgs=\ergscm. Low signal-to-noise measurements ($<3$) are reported as <$3\sigma$.}

    \label{tab:host_emission_lines}
\end{table}

\begin{table}[]
    \tiny
    \caption{List of emission lines from the AGN component, excluding \feii\ lines.}\label{tab:agn_emission_lines}

    \begin{tabularx}{\columnwidth}{Xll}
    \toprule
    Line & $v_{\rm FWHM}$ & Flux\\
    (1) & (2) & (3)\\
     & km/s & $10^{-18}$ cgs\\
    \midrule
    \multicolumn{3}{c}{Narrow lines ($z=6.63994\pm0.00003$)} \\
    \singlet{\oii}{3727.09} & $416.81\pm0.00$ & $2.76\pm0.45$ \\
    \singlet{\oii}{3729.88} & $416.50\pm1.90$ & $<1.23$ \\
    \Neiii{3869.86} & $401.37\pm1.83$ & $3.84\pm0.23$ \\
    \Hei{3889.75} & $399.31\pm1.82$ & $0.87\pm0.20$ \\
    \Neiii{3968.59} & $391.34\pm1.79$ & $1.43\pm0.26$ \\
    \singlet{\he}{3971.20} & $391.08\pm0.00$ & $<0.78$ \\
    \Siis{4069.75} & $381.57\pm1.74$ & $<0.45$ \\
    \Siis{4077.50} & $380.84\pm1.74$ & $<0.45$ \\
    \singlet{\hd}{4102.89} & $378.47\pm3.46$ & $1.24\pm0.23$ \\
    \singlet{\hg}{4341.68} & $357.54\pm0.00$ & $1.99\pm0.22$ \\
    \singlet{\oiii}{4364.44} & $355.66\pm3.25$ & $1.79\pm0.15$ \\
    \Hei{4472.735} & $347.00\pm1.59$ & $<0.33$ \\
    \Ariv{4713.57} & $329.14\pm3.01$ & $<0.35$ \\
    \Ariv{4741.49} & $327.19\pm1.50$ & $<0.34$ \\
    \singlet{\hb}{4862.68} & $318.97\pm1.46$ & $3.61\pm0.22$ \\
    \Hei{4923.304} & $315.01\pm1.44$ & $<0.36$ \\
    \singlet{\oiii}{4960.30} & $312.64\pm1.43$ & $10.85\pm0.13$ \\
    \singlet{\oiii}{5008.24} & $309.62\pm2.83$ & $32.33\pm0.46$ \\
    \Hei{5017.42} & $309.05\pm2.83$ & $0.65\pm0.12$ \\
    \singlet{\nii}{5756.24} & $268.99\pm2.46$ & $<0.37$ \\
    \Hei{5877.249} & $263.38\pm2.41$ & $1.15\pm0.14$ \\
    \singlet{\nai}{5891.58} & $262.73\pm3.61$ & $>-0.35$ \\
    \singlet{\nai}{5897.56} & $262.46\pm2.40$ & $>-0.38$ \\
    \Oi{6302.05} & $245.38\pm3.38$ & $<0.40$ \\
    \singlet{\nii}{6549.86} & $235.95\pm3.25$ & $<0.27$ \\
    \singlet{\ha}{6564.61} & $235.41\pm3.24$ & $9.88\pm0.58$ \\
    \singlet{\nii}{6585.27} & $234.66\pm2.15$ & $<0.79$ \\
    \Hei{6679.995} & $231.27\pm3.18$ & $0.54\pm0.18$ \\
    \singlet{\sii}{6718.29} & $229.93\pm3.17$ & $<0.51$ \\
    \singlet{\sii}{6732.67} & $229.43\pm3.16$ & $<0.50$ \\
     & & \\
    \multicolumn{3}{c}{BLR ($z=6.63994\pm0.00003$)$^\dagger$} \\
    \singlet{\hd}{4102.89} & $949.30\pm18.10$ & $<3.65$ \\
    \singlet{\hg}{4341.68} & $949.30\pm18.10$ & $1.81\pm0.47$ \\
    \singlet{\hb}{4862.68} & $949.30\pm18.10$ & $5.45\pm0.46$ \\
    \singlet{\ha}{6564.61} & $949.30\pm18.10$ & $89.12\pm2.51$ \\
    \bottomrule
    \end{tabularx}

    \tablefoot{(1) Line name including the vacuum wavelength (2) Observed FWHM velocity. (3) Observed line flux in cgs=\ergscm. Low signal-to-noise measurements ($<3$) are reported as <$3\sigma$. $^\dagger$ the BLR is exponential, therefore the FWHM is measured differently with $v_{\rm FWHM} = 2s\ln2c\lambda^{-1}$. No Gaussian outflow was measured for the AGN, due to the complex profile of oxygen and the good fit of the exponential BLR in Balmer lines. Therefore, as oxygen lines display weak wings in Fig.~\ref{fig:agn_spec_fig} due to contamination from the host, our narrow line estimation of their flux might be slightly over-estimated.}
\end{table}

\begin{table}[]
    \centering
    \tiny
    \caption{\feii\ emission line forest in the AGN component}
    
    \begin{tabularx}{\columnwidth}{Xll}
    \toprule
    Line & $v_{\rm FWHM}$ & Flux\\
    (1) & (2) & (3)\\
     & km/s & $10^{-18}$ cgs\\
    \midrule
    \Feii{4178.38} & $371.59\pm1.70$ & $<0.45$ \\
    \Feii{4246.01} & $365.64\pm1.67$ & $0.43\pm0.13$ \\
    \Feii{4278.04} & $362.89\pm1.66$ & $<0.40$ \\
    \Feii{4288.60} & $361.99\pm3.31$ & $0.46\pm0.13$ \\
    \Feii{4307.11} & $360.43\pm1.65$ & $<0.35$ \\
    \Feii{4320.84} & $359.27\pm1.64$ & $<0.38$ \\
    \Feii{4359.59} & $356.06\pm1.63$ & $<0.41$ \\
    \Feii{4415.02} & $351.56\pm1.61$ & $0.43\pm0.13$ \\
    \Feii{4815.89} & $322.10\pm2.94$ & $<0.36$ \\
    \Feii{4890.99} & $317.11\pm2.90$ & $<0.35$ \\
    \Feii{4906.72} & $316.09\pm1.45$ & $<0.35$ \\
    \Feii{5044.93} & $307.35\pm1.41$ & $<0.32$ \\
    \Feii{5160.23} & $300.42\pm2.75$ & $0.45\pm0.12$ \\
    \Feii{5263.10} & $294.49\pm1.35$ & $<0.33$ \\
    \Feii{5274.83} & $293.83\pm2.69$ & $<0.34$ \\
    \bottomrule
    \end{tabularx}

    \tablefoot{(1) Emission lines with associated vacuum wavelength (2) Observed FWHM velocity (3) Flux in cgs = \ergscm. Low signal-to-noise measurements ($<3$) are reported as <$3\sigma$.}
    
    \label{tab:agn_iron_forest}
\end{table}

\section{Comparison between the nod and global sky background subtraction}\label{app:nod_vs_gsky}

In Sect.~\ref{sec:data}, we mentioned that due to over-subtraction of the sky background with the nod reduction, due to the galaxy being larger than the NIRSpec slit, we used an alternative reduction based on a global estimation of the sky background \citep[as introduced in the RUBIES survey paper][]{degraaff_rubies_2025}. We use this global sky background subtraction for all our main measurements, but use the nod reduction in the measurements requiring the continuum: Figs.~\ref{fig:balmer_break} and \ref{fig:black_body}, and the estimation of $\beta_{\rm opt}$ in the discussion. These are all secondary analyses and independent from the main results.

\begin{figure*}
\centering
\includegraphics[width=0.47\textwidth]{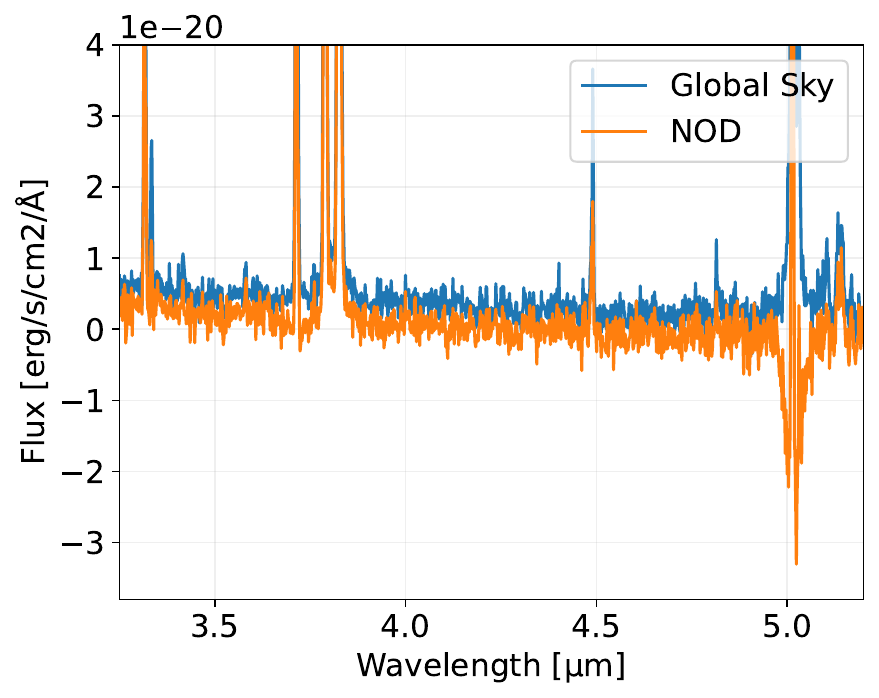}
\includegraphics[width=0.49\textwidth]{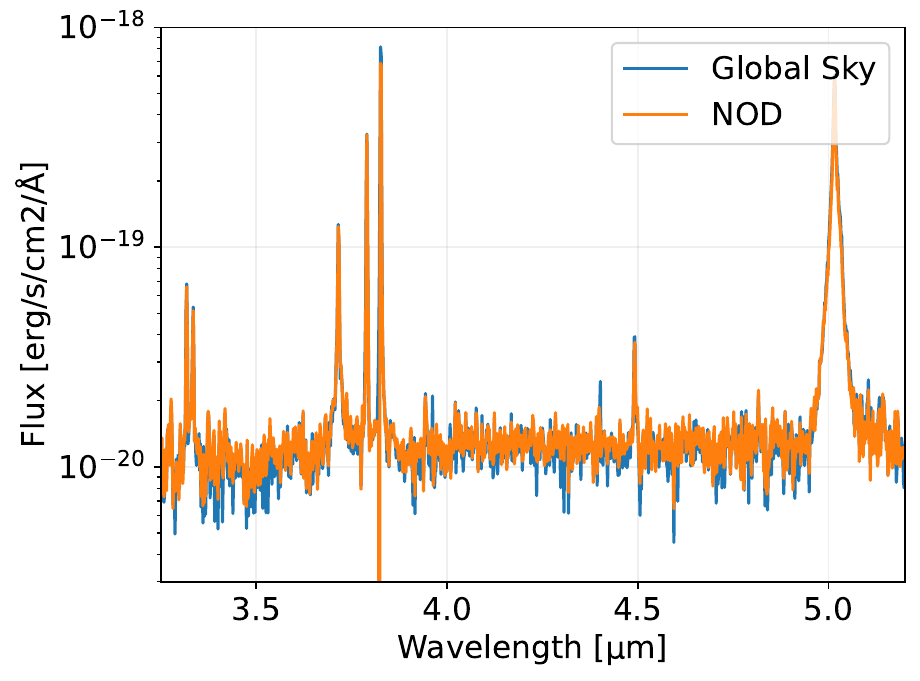}
\caption{Comparison between the nod sky background subtraction (orange) and the global sky background subtraction (blue) for the host (left) and the AGN (right) in the region of interest.}
\label{fig:nod_vs_gsky}
\end{figure*}

In Fig.~\ref{fig:nod_vs_gsky}, we show the difference between these two subtractions for the AGN and the host spectra. In the AGN (right), more or less no difference is observed, due to the fact that the AGN does not lie in the over-subtracted region of the 2D spectrum. However, the over-subtraction can be observed in the host (left), where the continuum and the emission lines in the nod subtraction are lower, and the wings of \ha\ and \oiii\ are heavily affected. We never use the host nod continuum in our measurements, but we do the few measurements mentioned in the previous paragraph on the AGN nod continuum. 

\section{Statistical analysis of the outflow}\label{app:stat_ana_outflow}

\begin{figure}
    \centering
    \includegraphics[width=1\linewidth]{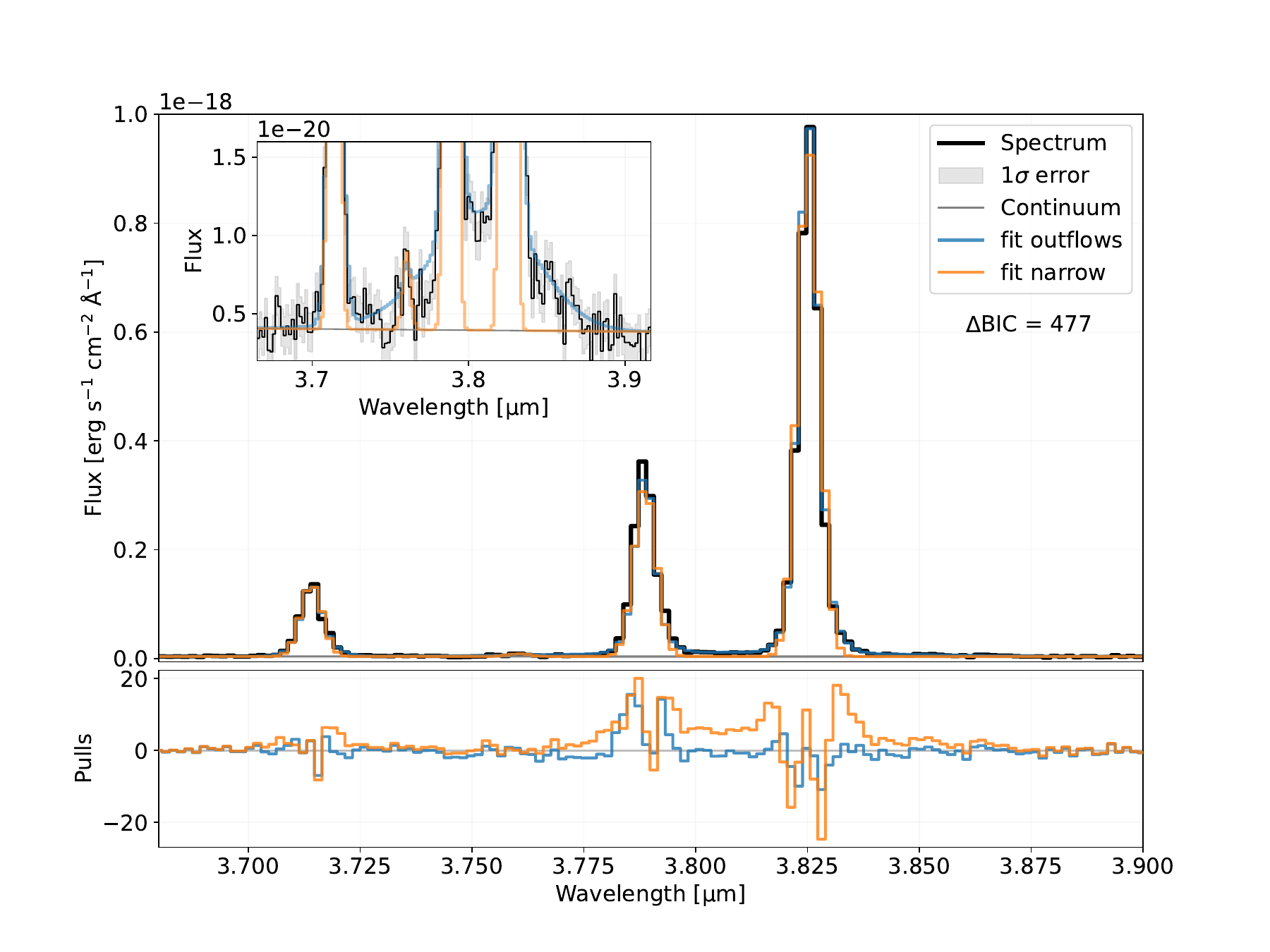}
    \caption{Visual comparison of the difference, centred around \Oiiihb,  between the fit including the two outflowing (blue) and with only narrow components lines (orange). Top panel: The data (black) and the two fits are displayed. The main panel shows the full lines, emphasising on the high signal-to-noise regions (peaks) and a zoom-in region centred on the outflow shows the differences in the low signal-to-noise regions (around the outflow). The units of the flux in the zoom-in are the same as the main panel. Bottom panel: The pull = (data - model)/uncertainty of both fits.}
    \label{fig:outflow_vs_no}
\end{figure}

In this appendix, we statistically analyse the difference between the fit with and without outflowing components. In Fig.~\ref{fig:outflow_vs_no} we show this difference centred around \Oiiihb. We only display \oiiihb, due to these lines being the strongest lines affected by the outflow, but the fit is applied on all the emission lines of interest. All measurements in this section are then performed on the whole spectrum (see Sect.~\ref{sec:fit_host}). The fit with outflowing components (blue) is presented in the rest of the paper, while the narrow fit (orange) removes all outflowing components in the host. We observe that the fit with outflowing components generally reproduces the data much better, as also confirmed by the much better BIC ($\Delta{\rm BIC}=477$, measured on the global fit). Nonetheless, we still observe differences in the Pull due to the hard constraints on the \oiii\ line ratios, and the non-Gaussian behaviour of the outflows. To better understand the outflow in \oiii, we need more advanced physical models that account better for the physical properties and spatial information of the outflow. However, the limited information about this object restricts our ability to model this outflow with more advanced models. 

\section{Continuum of the AGN}\label{app:agn_continuum}

\begin{figure}
    \centering
    \includegraphics[width=1\linewidth]{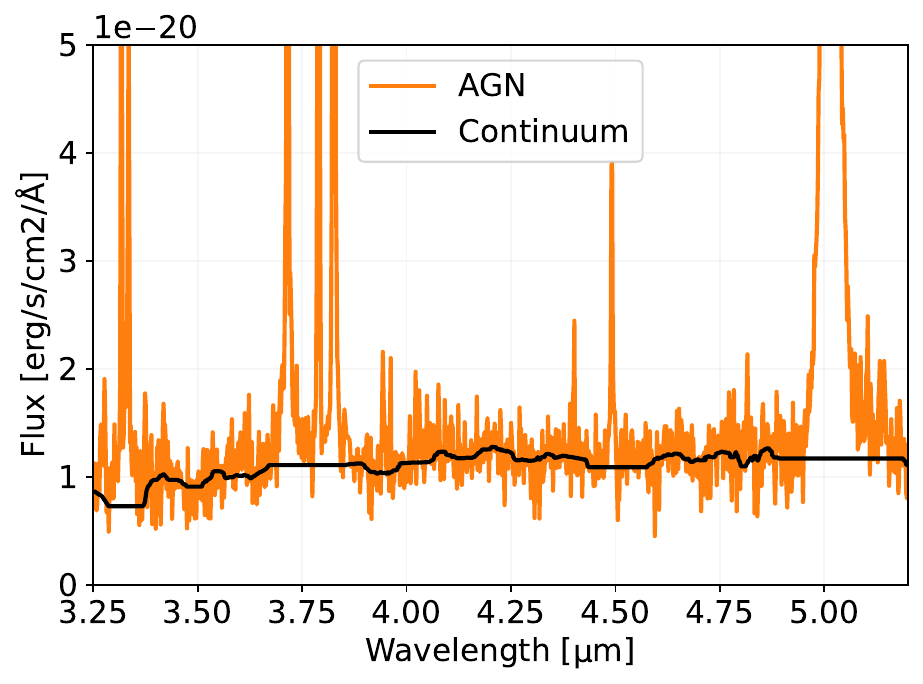}
    \caption{Fit of the AGN continuum in the stable region as described Sect.~\ref{sec:fit_agn}.}
    \label{fig:agn_continuum}
\end{figure}

In Fig.~\ref{fig:agn_continuum}, we display the fitted continuum to extract the emission-line fluxes. Most regions are fitted automatically by discarding a small window around emission lines, but the continuum in the strong and large emissions was manually fixed (see Sect.~\ref{sec:fit_agn}).

\section{Table of maximum velocity}\label{app:vout}

There exist multiple definitions of the maximum velocity $v_{\rm out}$ in the literature. Assuming that $\Delta v$ is the difference in velocity between the narrow and outflow peaks, and $\sigma_{\rm out} = v_{\rm FWHM}/2.355$ is the standard deviation of the outflow Gaussian component: $v_{\rm out} = \abs{\Delta v} + v_{\rm FWHM}/2$ \citep[i.e. $\sim90\%$ of the gas has a velocity lower than $v_{\rm out}$, e.g.][]{Rupke_outflows_2005, arribas_ionized_2014}, $v_{\rm out} = \abs{\Delta v} + 2\sigma_{\rm out}$ \citep[i.e. $\sim98\%$ of the gas has a velocity lower than $v_{\rm out}$, e.g.][]{Genzel_sins_2011, Rupke_multiphase_2013} or others \citep[e.g.][]{forster_schreiber_kmos3d_2019, bertola_ga-nifs_2025}. In this paper we used the definition from \citet{Genzel_sins_2011} and \citet{Rupke_multiphase_2013} and we harmonised all the results to this maximum velocity. In Tab.~\ref{tab:vout}, we report the measurements of $v_{\rm out}$ and total (if possible) \ha\ luminosity of multiple star-formation and AGN-driven outflows and known outflowing LRDs. These values were already displayed in Fig.~\ref{fig:vout}.

\begin{table}[]
    \centering
    \caption{Tabulated values as shown in Fig.~\ref{fig:vout}.}
    \tiny
    \begin{tabularx}{\columnwidth}{Xlll}
    \toprule
    Target & $z$ & $v_{\rm out}$ & $L_{\ha}$\\
    (1) & (2) & (3) & (4)\\
    & & \kms & $10^{42}$\ergs\\
    \midrule
    \multicolumn{4}{c}{Little red dot (LRD)} \\
    \multicolumn{4}{c}{\citet{deugenio_jades_2026}}\\
    159717 & $5.078$ & $269.00 \pm 41.48$ & $7.94 \pm 0.13$ \\
    \multicolumn{4}{c}{\citet{cooper_high-velocity_2025}}\\
    2389 & $7.043$ & $757.71 \pm 163.65$ & $0.99 \pm 0.00$ \\
    \multicolumn{4}{c}{\citet{chen_z_2026}}\\
    J2048 & $0.433$ & $3097.39 \pm 2219.89$ & $10.16 \pm 0.16$ \\
    \multicolumn{4}{c}{This work}\\
    GL-329380 & $6.638$ & $5137.01 \pm 499.04$ & $16.62 \pm 0.96$ \\
    &&&\\
    \multicolumn{4}{c}{Active Galaxy Nucleus (AGN)} \\
    \multicolumn{4}{c}{\citet{bertola_ga-nifs_2025}}\\
    COS590 & $3.52385$ & $796.00 \pm 5.00$ & $1.21 \pm 0.02$ \\
    COS1118 & $3.643$ & $823.00 \pm 3.00$ & $0.63 \pm 0.02$ \\
    COS349 & $3.5093$ & $1384.00 \pm 8.00$ & $0.79 \pm 0.04$ \\
    COS1656-A & $3.5101$ & $1326.00 \pm 6.00$ & $1.84 \pm 0.16$ \\
    COS2949 & $2.0478$ & $901.00 \pm 3.00$ & $0.57 \pm 0.01$ \\
    COS1638-A & $3.5057$ & $3551.00 \pm 131.00$ & $3.39 \pm 0.03$ \\
    COS1638-B & $3.5109$ & $2623.00 \pm 56.00$ & $30.73 \pm 0.09$ \\
    \multicolumn{4}{c}{\citet{ubler_ga-nifs_2023}}\\
    GS\_3073 & $5.55$ & $1211.00 \pm 20.00$ & $33.09 \pm 0.55$ \\
    &&&\\
    \multicolumn{4}{c}{Star formation (SF)} \\
    \multicolumn{4}{c}{\citet{saldana-lopez_feedback_2025}}\\
    1871-10 & $7.59903$ & $301.60 \pm 50.80$ & $2.76 \pm 0.56$ \\
    1871-12 & $7.50201$ & $360.32 \pm 39.06$ & $6.10 \pm 0.77$ \\
    1871-14 & $7.20485$ & $293.11 \pm 84.18$ & $1.58 \pm 0.31$ \\
    1871-40 & $7.09481$ & $523.23 \pm 90.20$ & $1.30 \pm 0.35$ \\
    1871-105 & $6.56692$ & $252.85 \pm 30.64$ & $1.18 \pm 0.13$ \\
    \multicolumn{4}{c}{\citet{llerena_ionized_2023}}\\
    CDFS019946 & $2.437$ & $183.13 \pm 27.91$ & $2.02 \pm 0.23$ \\
    CDFS018182 & $2.317$ & $253.91 \pm 10.76$ & $5.77 \pm 0.33$ \\
    CDFS022799 & $2.544$ & $223.58 \pm 42.42$ & $8.67 \pm 0.83$ \\
    UDS021398 & $2.492$ & $307.31 \pm 23.59$ & $5.73 \pm 0.42$ \\
    5100998761 & $2.453$ & $302.39 \pm 21.07$ & $2.72 \pm 0.49$ \\
    5101421970 & $2.47$ & $238.89 \pm 32.90$ & $5.72 \pm 0.60$ \\
    530048433 & $2.312$ & $337.02 \pm 11.86$ & $7.01 \pm 0.23$ \\
    510583858 & $2.417$ & $328.75 \pm 17.07$ & $9.76 \pm 0.99$ \\
    511001501 & $2.227$ & $193.53 \pm 22.26$ & $7.59 \pm 0.74$ \\
    \bottomrule
    \end{tabularx}
    \label{tab:vout}
    \tablefoot{(1) Target name (2) Systemic redshift (3) Maximum velocity harmonised to the $v_{\rm out}=\abs{\Delta v} + 2\sigma_{\rm out}$ convention (4) The total $\ha$ luminosity. \citet{cooper_high-velocity_2025} only provides the measurements of $v_{\rm out}$ for their LRD. The $L_{\ha}$ is inferred from $2.86\times L_{\hb}$ following the values of \citet{rinaldi_not_2025}. However, the flux of \hb\ is not clearly defined as broad or narrow. \citet{bertola_ga-nifs_2025} only provides the flux of the broad component; therefore the real $L_{\ha}$ is expected to be stronger. \citet{deugenio_jades_2026} provides asymmetric uncertainty measurements, so we approximated all uncertainties with a symmetric Gaussian.}
\end{table}

\end{document}